\documentclass[conference]{IEEEtran}
\IEEEoverridecommandlockouts
\usepackage{cite}
\usepackage{amsmath,amssymb,amsfonts}
\usepackage{algorithmic}
\usepackage{graphicx}
\usepackage{textcomp}
\usepackage{xcolor}
\usepackage{subcaption}
\usepackage{hyperref}
\usepackage{wrapfig}
\usepackage{fancyhdr}
\def\BibTeX{{\rm B\kern-.05em{\sc i\kern-.025em b}\kern-.08em
    T\kern-.1667em\lower.7ex\hbox{E}\kern-.125emX}}

\begin{document}

%\title{Micro-Architectural Bottlenecks in LLM Inference: Quantifying the Impact of SRAM, Frequency, and Memory Bandwidth on Energy Efficiency\\}

\title{Prefill vs. Decode Bottlenecks: SRAM–Frequency Tradeoffs and the Memory-Bandwidth Ceiling}

\author{
\IEEEauthorblockN{Hannah Atmer}
\IEEEauthorblockA{
Uppsala University\\
Sweden\\
0000-0002-7033-6373
}
\and
\IEEEauthorblockN{Yuan Yao}
\IEEEauthorblockA{
Uppsala University\\
Sweden\\
0000-0001-9448-5595
}
\and
\IEEEauthorblockN{Thiemo Voigt}
\IEEEauthorblockA{
Uppsala University\\
Sweden\\
0000-0002-2586-8573
}
\and
\IEEEauthorblockN{Stefanos Kaxiras}
\IEEEauthorblockA{
Uppsala University\\
Sweden\\
0000-0001-8267-0232
}
}

\maketitle
\begin{abstract}

Energy consumption dictates the cost and environmental impact of deploying Large Language Models. This paper investigates the impact of on-chip SRAM size and operating frequency on the energy efficiency and performance of LLM inference, focusing on the distinct behaviors of the compute-bound prefill and memory-bound decode phases. Our simulation methodology combines OpenRAM for energy modeling, LLMCompass for latency simulation, and ScaleSIM for systolic array operational intensity. Our findings show that total energy use is predominantly determined by SRAM size in both phases, with larger buffers significantly increasing static energy due to leakage, which is not offset by corresponding latency benefits. We quantitatively explore the memory-bandwidth bottleneck, demonstrating that while high operating frequencies reduce prefill latency, their positive impact on memory-bound decode latency is capped by the external memory bandwidth. Counter-intuitively, high compute frequency can reduce total energy by reducing execution time and consequently decreasing static energy consumption more than the resulting dynamic power increase. We identify an optimal hardware configuration for the simulated workload: high operating frequencies (1200MHz-1400MHz) and a small local buffer size of 32KB to 64KB. This combination achieves the best energy-delay product, balancing low latency with high energy efficiency. Furthermore, we demonstrate how memory bandwidth acts as a performance ceiling, and that increasing compute frequency only yields performance gains up to the point where the workload becomes memory-bound. This analysis provides concrete architectural insights for designing energy-efficient LLM accelerators, especially for datacenters aiming to minimize their energy overhead.
\end{abstract}


\thanks{This work was supported by the Swedish Foundation for Strategic Research (SSF) grant FUS21-0067.}

\section{Introduction}
Data centers, which host LLMs and other computational infrastructure, globally consumed an estimated 415 Terawatt-hours (TWh) of electricity in 2024, accounting for about 1.5\% of worldwide electricity demand, with projections indicating this consumption could more than double to around 945 TWh by 2030 due largely to the growth of energy-intensive AI workloads like LLMs~\cite{iea}. Energy consumption during LLM inference is an important metric for assessing the cost, scalability, and environmental impact of deploying LLMs~\cite{rado, efficient}. LLM inference comprises two distinct phases with different architectural bottlenecks. The \textit{prefill} phase, which processes the entire input prompt, involves massive matrix multiplications that are highly parallelizable, making it \textit{compute-bound}. In contrast, the subsequent \textit{decode} phase, which generates output token-by-token, involves running the full model for each new token. Despite the relatively smaller amount of computation per step, the necessity of repeatedly streaming billions of model parameters from off-chip memory makes the decode phase \textit{memory-bound}.

%  basics of static vs dynamic energy
Energy consumption in GPUs consists of static and dynamic energy consumption. Static energy is consumed continuously even when no dynamic power is expended. Power gating, especially in SRAM~\cite{kaxiras2001cache,flautner2002drowsy}, can reduce static energy and we account for such techniques~\cite{npu} in our study. Dynamic energy, on the other hand, is consumed proportionally to activity. If a GPU is idle then it consumes very little dynamic power, assuming proper clock gating. Compute latency, the amount of work performed per unit of time, and the number of transistors on the chip, will affect the amount of static and dynamic energy consumption of LLM inference.

% previous work to study energy use + we study effect of frequency vs sram size on energy consumption + main insight: why we look at frequency and on-chip SRAM
%Larger on-chip SRAM can decrease latency but increases static energy consumption, and higher frequency decreases compute latency but increases dynamic energy consumption.

While the memory-bound nature of decode is well-established, prior studies on LLM inference energy use~\cite{quantifying, ellie, MNN, efficient, hungry} do not consider the effect of on-chip SRAM size ($S$) and compute frequency ($f$). These parameters present an inherent trade-off that affects the total energy consumption. Specifically, increasing the on-chip SRAM size can decrease memory access latency but significantly increases static energy consumption. Increasing frequency reduces compute latency but increases dynamic energy consumption. The interaction between static (leakage) power (a function of execution time) and dynamic power (a function of activity and frequency) 
%\tv{I do not know the exact definitions but are activity and execution distinct concepts?}
is complex and must be quantified to determine true energy efficiency.

Therefore, this work quantitatively measures the effect of on-chip SRAM size and frequency on LLM energy consumption and performance during both prefill and decode, with a focus on understanding the memory-bound characteristics of decode because decode uses significantly more energy than prefill~\cite{MNN}. We further explore the performance limitations by examining the impact of varying memory bandwidth on latency and energy consumption, demonstrating how the memory bottleneck dictates the maximum effective compute frequency. Using a comprehensive simulation methodology combining OpenRAM~\cite{guthaus2016openram} for power modeling, LLMCompass~\cite{zhang2024llmcompass} for latency, and ScaleSIM~\cite{raj2025scalesim} for operational intensity, we investigate the architectural sweet spot. We demonstrate that there exists an optimal frequency range (1200MHz–1400MHz) and local buffer size (32KB–64KB) that effectively minimizes the Energy-Delay Product (EDP) by balancing static leakage costs against performance gains.

Decreasing LLM inference energy consumption is important for the widespread societal integration of AI because it determines the environmental sustainability~\cite{hungry} and economic viability of deploying LLMs. Therefore, understanding how LLM inference energy consumption is affected by parameters like SRAM size and frequency can provide insight about which chip specifications and compute frequencies will minimize an LLM deployment's energy cost.

The remainder of the paper is organized as follows. Section II provides background on on-chip SRAM and LLM inference phases. Section III details the simulation method, including OpenRAM, LLMCompass, and ScaleSIM. Section IV presents the results on cycles, latency, and energy metrics. Finally, Section V concludes the work and suggests directions for future research.

\section{Background}

This section provides an overview of 1. on-chip SRAM, 2. LLM inference, and 3. systolic arrays.

\subsection{On-Chip SRAM}

%SRAM (Static Random-Access Memory) is volatile memory that stores data using bistable latching circuitry (flip-flops), meaning it retains its content as long as power is supplied without needing to be refreshed like DRAM. SRAM has nanosecond response times and is often used for caches and on-chip buffers. 
In GPUs or custom accelerators that use systolic arrays (e.g. Google’s TPU or ML-specific ASICs), SRAM plays a central role in performance, energy efficiency, and dataflow optimization. In systolic array-based accelerators, SRAM holds inputs, weights, and partial results, enabling high reuse, low latency, and efficient execution of operations like matrix multiplications which are fundamental in deep learning workloads\cite{tpu}. In a Systolic Array-based GPU Architecture, SRAM acts as local scratchpad memory for processing elements: it stores weights, input activations, and partial results~\cite{purdue}. Data is fetched from DRAM in chunks and held in SRAM for low-latency, high-throughput access by the compute array.

\subsection{LLM Inference}

The prefill phase initiates LLM inference by processing the entire input prompt. This phase employs the Transformer's self-attention mechanism to compute the relationships between all input tokens, generating all necessary Key and Value (KV) tensors and building the KV cache. This process involves immense matrix multiplications that are highly suitable for a GPU’s parallel architecture. Because prefill is significantly more computationally expensive per token than decoding and exhibits quadratic complexity with respect to sequence length, this phase is generally classified as compute-bound. The decode phase generates the LLM output auto-regressively, predicting one token at a time and adding it back to the sequence. Instead of recomputing attention over the entire history, the model utilizes the Key-Value (KV) cache built during prefill, allowing the new token to attend only to the cached key/value vectors. However, each step requires streaming the entire model's billions of parameters from memory into the cache. This massive data movement, combined with relatively small per-step computation, makes the decode phase inherently memory-bound. Decode dominates the overall LLM inference energy consumption due to its much longer execution time~\cite{MNN}.

\subsubsection{Systolic Arrays}
In a systolic array, data (activations, partial sums, or weights) flows through a grid of processing elements (PEs). Systolic arrays are often weight-stationary. In a weight-stationary design, weights are loaded into the PEs once and held stationary. Activations (input vectors/matrices) stream through the array. This minimizes weight re-reads from external DRAM, reducing bandwidth pressure. While stationary weights improve data reuse within a single matmul, the need to reload enormous weight matrices for each token means decode remains memory-bandwidth bound.

\section{Method}

We use OpenRAM~\cite{guthaus2016openram} to calculate static and dynamic energy consumption. We assume savings in static energy by 4\% for prefill and 20\% for decode based on ideal power gating results found by Xue and Huang~\cite{npu}. We use LLMCompass~\cite{zhang2024llmcompass} to simulate the latency of the matrix multiplications of an LLM runtime, for each frequency. And finally, we estimate the number of ops per cycle using ScaleSim~\cite{raj2025scalesim}.

\begin{figure}[t]
    \centering
    \includegraphics[width=\linewidth]{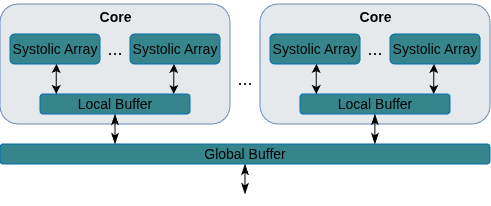}
    \caption{Modeled architecture: We assume a generic accelerator with a global buffer that inrefaces to memory and local buffers that serve the systolic arrays.}
    \label{fig:hw_model}
\end{figure}

The hardware modeled in our experiments, simplified in \autoref{fig:hw_model}, is an accelerator featuring 108 processing cores. Each core contains four $16 \times 16$ systolic arrays and one local buffer. These cores share a 40 MB global buffer. Our model accounts for the memory bandwidth between the local buffers and the global buffer, as well as the bandwidth of data transfers into and out of the global buffer. External memory is not included in our model.

\begin{figure*}[ht]
    \centering
    \includegraphics[width=\linewidth]{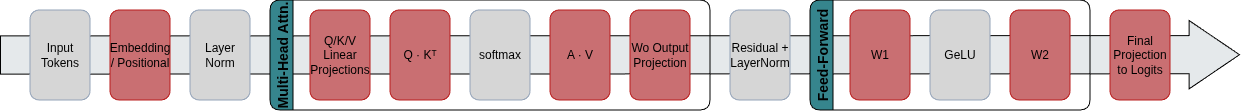}
    \caption{LLMCompass computation and data transfer taken into account. Omitted layers contribute a negligible amount to either computation or data tranfer.}
    \label{fig:layers}
\end{figure*}

\subsection{LLMCompass}
We use a modified version of LLMCompass to simulate how long it takes to perform the matrix multiplications of LLM prefill and decode stages. In the prefill stage, we record the latency for processing the entire input prompt, and for the decode stage we record the latency for generating a single token of the output. 
We measure only the latency of the layers that predominantly perform matrix multiplications because this makes the results more generalizable to many different neural network deployments. \autoref{fig:layers} shows the matmul-heavy layers of the simulated neural network in red; LLM layers that are not modeled contribute relatively negligible amounts of computation and data transfer, compared to the matmul-heavy layers. 

In transformer-based LLMs, QKV projection, Attention score $(Q·K^T)$, Attention output $(weights·V)$, MLP up-projection, and MLP down-projection dominate compute. These sublayers are present in every transformer block, and transformer blocks are used in both prefill and decode. The layers are the same in both phases, but the sequence length changes the dominant cost. In prefill, attention matmuls are expensive because $(Q·K^T)$ and $(weights·V)$ grow with context length. In decode, attention matmuls shrink dramatically (only 1 token as Q), but MLP matmuls dominate because they are independent of sequence length. In the decode phase, the key-value (KV) cache grows with each generated token. LLMCompass assumed a fixed-size KV cache, so we modified the LLMCompass code to simulate the autoregressive generation of tokens and the associated growth of the KV cache. This improved the accuracy of the performance model for evaluating the transformer's performance during inference. The "prompt" LLMCompass uses for its latency measurements is an input batch of size 8, each with a sequence length of 2048 tokens, processed through one layer of GPT-3~\cite{zhang2024llmcompass}. Decode latency is measured per output token~\cite{zhang2024llmcompass}.

\subsection{Configurations}

We use LLMCompass to measure the latency of a simulated generic accelerator with 128 16x16 systolic arrays. We model the accelerator at a 45nm process node. We model SRAM arrays (local and global buffers) with OpenRAM~\cite{guthaus2016openram} to estimate static and dynamic energy use for a range of SRAM sizes.  Specifically, we use OpenRAM to derive leakage power and energy per clock edge, from which we can deduce energy per cycle, for each SRAM size. We use ScaleSIM to find the number of memory and compute ops per cycle. Specifically, we get average SRAM reads and writes per cycle so that we can calculate dynamic energy, as well as the systolic array's average utilization.

For the local buffers, we do a sweep from 16KB to 256KB, in contrast to Shang et al. who only looked at larger local buffers but with little benefit over 256KB~\cite{zhang2024llmcompass}. The job of the global buffer is to provide a staging area between the local buffers and systolic arrays and memory. Specifically, this staging area decouples the local buffers and systolic arrays from memory by double-buffering loads from memory and providing ample and fast storage for storing the output of the systolic arrays before it is written out to memory. This decoupling allows the systolic arrays and their local buffers to operate at peak performance, once they are loaded with the necessary data. What we observe in our studies is that the size of the global buffer does not significantly affect performance. The same result is corroborated by Zhang et al.~\cite{zhang2024llmcompass}: Quadrupling the size of the global buffer from 10MB to 40MB, increases prefill performance by 11\% but doubling the size again to 80MBsonly yields a negligible 0.01\% performance increase~\cite{zhang2024llmcompass}; similarly, increasing the global buffer size by 8$\times$ yields (10MB to 80MB) in decode, only yields a minuscule 0.7\% performance increase~\cite{zhang2024llmcompass}. This reflects our own experience, irrespective of the other parameters we change. For this reason, we assume a standard 40MB global buffer for all cases, which we model for energy (dynamic and static), but we do not perform additional experiments with other sizes. Finally, we note that by varying memory bandwidth, the global buffer size becomes more relevant as a parameter, but again the effects are not pronounced. 

\begin{wrapfigure}{r}{0.3\linewidth}
    \centering
    \vspace{-10pt}
    \includegraphics[width=\linewidth]{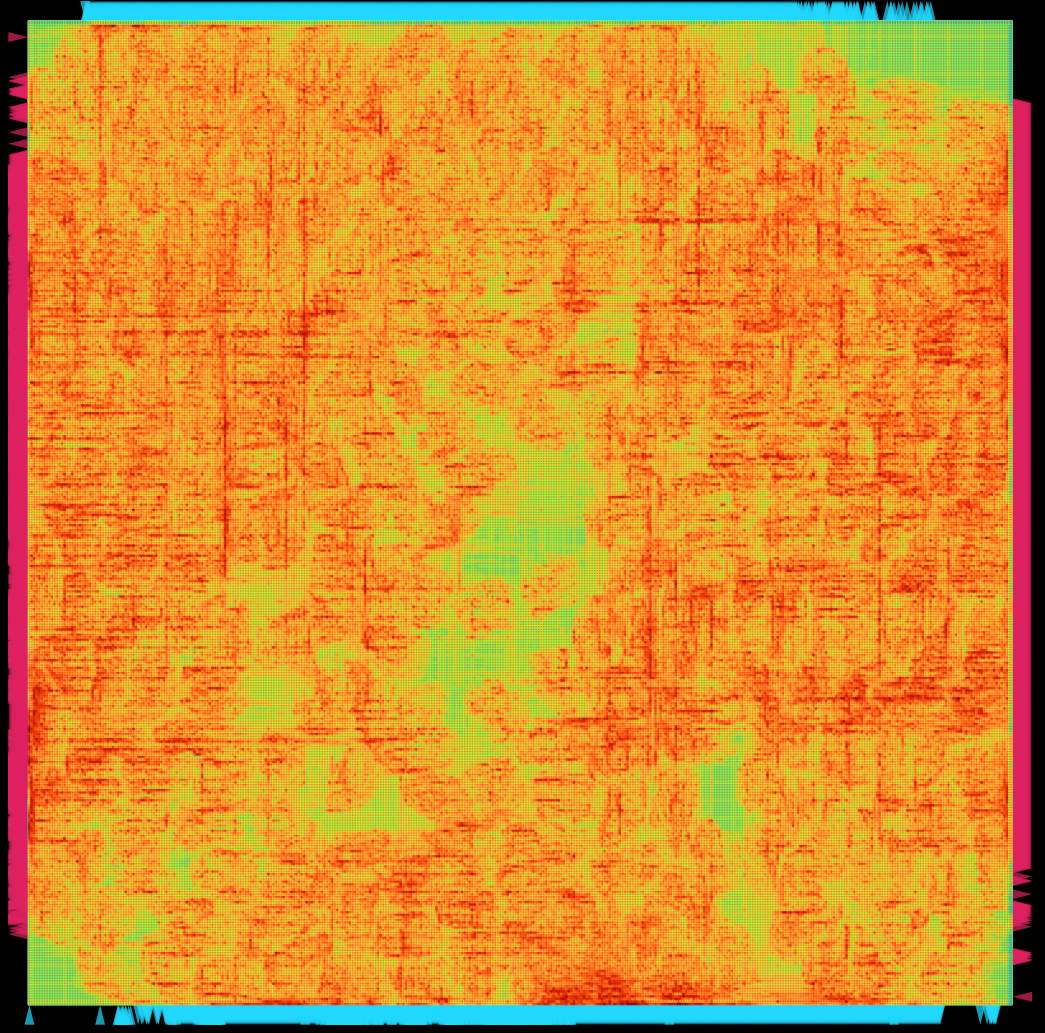}
    \vspace{-10pt}
    \caption{Physical design of the 16$\times16$ systolic array with routing-wire density}
    \label{fig:synth}
\end{wrapfigure}
To estimate the energy of the systolic arrays, we use Yosys~\cite{yosys} for logic synthesis and OpenROAD~\cite{openroad_1,openroad_2} for physical implementation. We target the NanGate45 standard-cell library, which is based on the FreePDK45 process~\cite{freepdk}. The backend flow follows floorplanning, placement using a 10-layer metal stack, clock-tree synthesis (CTS), and routing, resulting in a finalized layout with a core area of $6.60\times10^{5}\,\mu\mathrm{m}^2$ (as shown in Figure~\ref{fig:synth}), and an optimized standard-cell placement at 56.0\% area utilization. Post-layout power analysis reports a leakage (static) power of $9.31\,\mathrm{mW}$ and a dynamic power of $1.25\,\mathrm{W}$, including $0.59\,\mathrm{W}$ internal power due to charging/discharging within the cells and $0.65\,\mathrm{W}$ switching power dominated by interconnect activity and charging of routed net capacitances. In our default configuration used for the majority of our results, the accelerator uses a a 2048 GB/s memory bandwidth. In this work we study the on-chip resources and we do not model energy for memory accesses.

%\section{Results}
\section{Evaluation}

%%%%%%%% Latency, power, static and dynamic energy, and total energy %%%%%%%%%

\subsection{Architectural Performance (cycles)}

We begin our analysis from an architectural perspective. In particular, we examine how frequency $f$ and local buffer size $S$ affect the number of cycles it takes to execute a fixed amount of LLM work. Previous works have noted that the behavior of Prefill and Decode with respect to memory- or compute-\emph{boundness} differ dramatically~\cite{zhang2024llmcompass, dao2023flashattention2fasterattentionbetter}. More specifically, Prefill is \emph{compute-bound} and Decode is \emph{memory-bound}. This means that they are asymmetrically affected by $f$. In this work, we examine the relation of the bound-ness of each of the Prefill and Decode, not only as a function of $f$ but also in relation to the local buffer size $S$. Considering only $f$ in isolation, previous work has shown, both with analytical models~\cite{Keramidas,Lieven,Kaxiras_Martonosi_2008} and empirically~\cite{Spiliopoulos}, that the performance of \emph{compute-bound} workloads is directly affected by $f$ (higher-frequency, better performance in terms of latency or delay). In contrast, the performance of \emph{memory-bound} workloads is ``elastic'' to $f$, meaning that for significant stretches of $f$ performance remains largely unaffected~\cite{Keramidas,Lieven,Kaxiras_Martonosi_2008,Spiliopoulos}.

%%%% Cycle Count %%%%
\begin{figure}[ht]
    \centering
    \includegraphics[width=\linewidth]{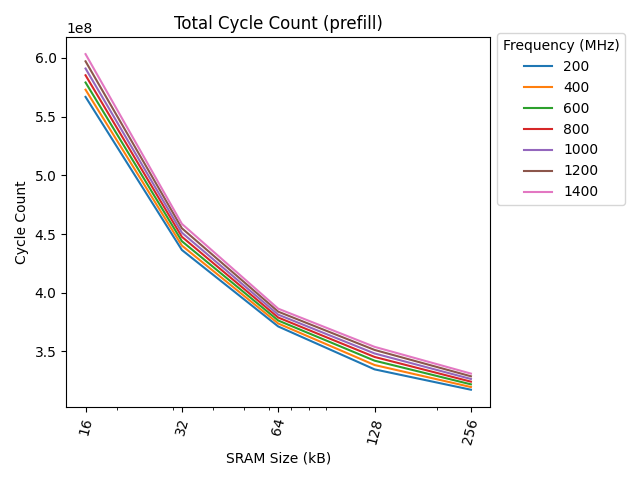}
    \caption{Prefill Cycle Count as a function of frequency (minor effect) and 1st level SRAM buffer (major effect)}
    \label{fig:prefill_cycle_count}
\end{figure}
\begin{figure}[ht]
    \centering
    \includegraphics[width=\linewidth]{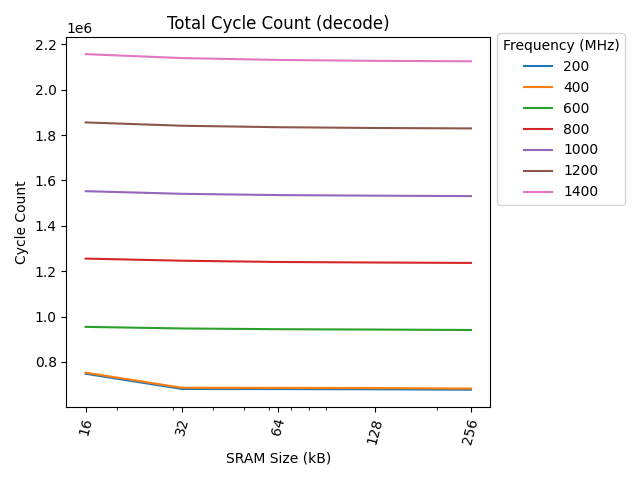}
    \caption{Decode Cycle Count as a function of frequency (major effect) and 1st level SRAM buffer (minor effect)}
    \label{fig:decode_cycle_count}
\end{figure}

Figures \ref{fig:prefill_cycle_count} and \ref{fig:decode_cycle_count} show how many cycles the LLM workload requires to complete for each $f$ and $S$, for prefill and decode respectively. Prefill cycle counts are only slightly affected by $f$ and decrease with increasing $S$. Decode cycle counts increase with increasing $f$, and are mostly unaffected by $S$. Since the memory operations (fetching the KV-cache and weights) take a fixed amount of real time (latency) due to the memory bus speed, shortening the clock cycle (increasing frequency) means that the fixed real time now spans more clock cycles. 

\begin{figure*}[ht]
  \centering

  \begin{subfigure}{0.48\textwidth}
    \centering
    \includegraphics[width=\linewidth]{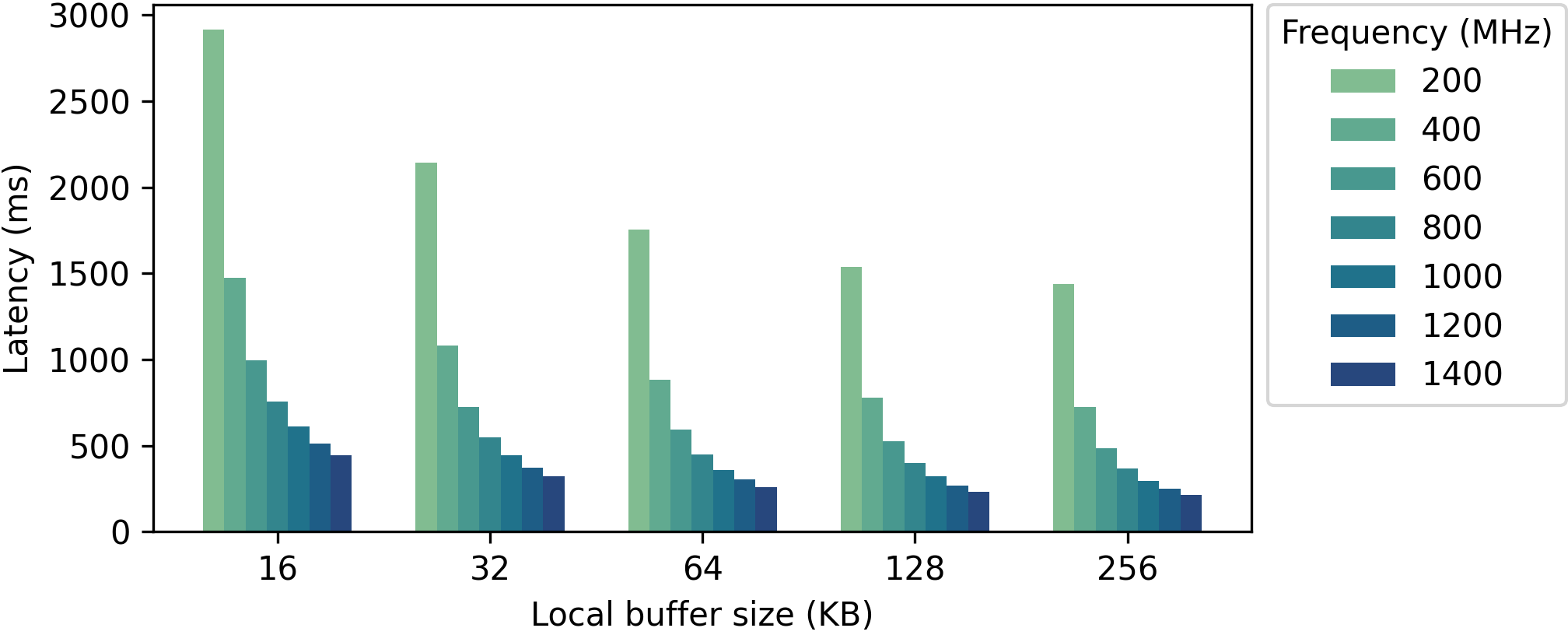}
    %\caption{Prefill Latency (to encode a full prompt) as a function of local buffer size (x-axis) and frequency (7 bars per x-axis point)}
    %\label{prefill_latency}
  \end{subfigure}
  \hfill
  \begin{subfigure}{0.48\textwidth}
    \centering
    \includegraphics[width=\linewidth]{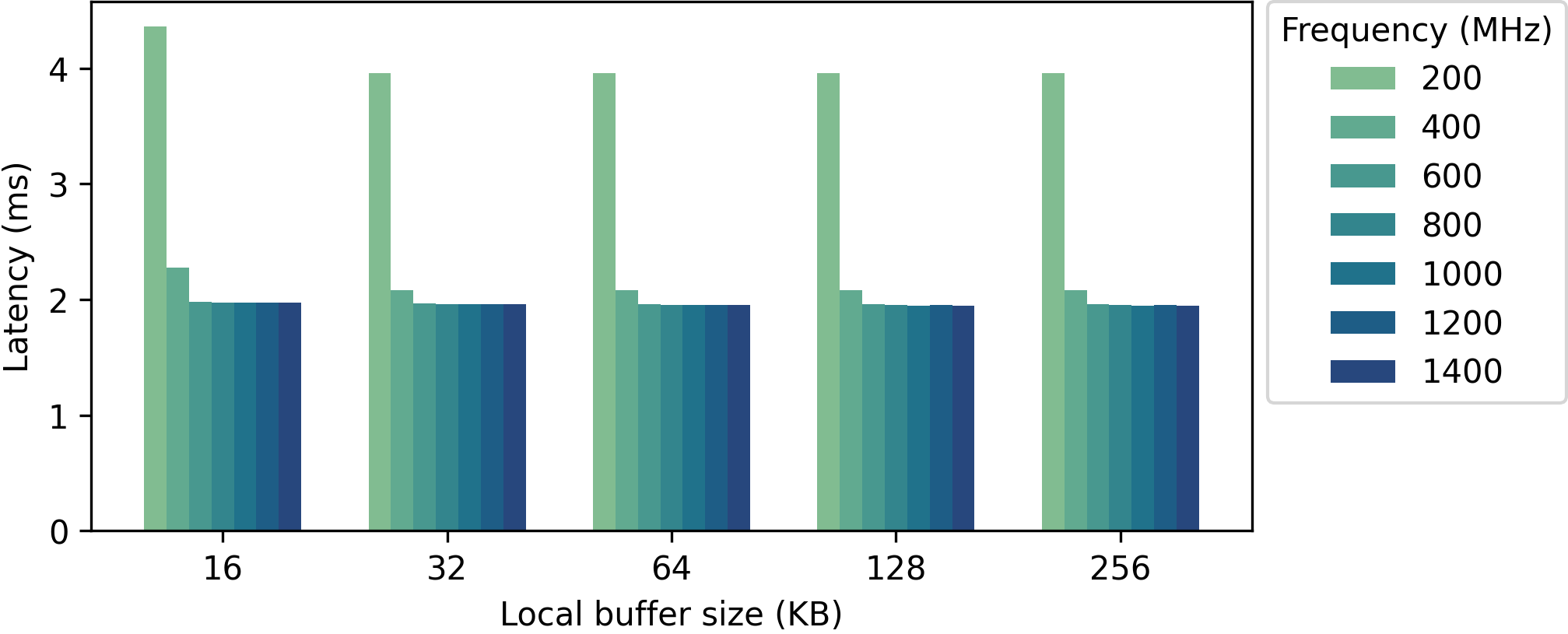}
    %\caption{Decode Latency per token as a function of local buffer size (x-axis) and frequency (7 bars per x-axis point). 2048 GB/s memory bandwidth.}
    %\label{decode_latency}
  \end{subfigure}
    \caption{Prefill latency, (a)--left graph, and decode latency, (b)--right graph,  as a function of local buffer size (x-axis) and frequency (7 bars per x-axis point), and constant 2048 GB/s memory bandwidth.
    %: (a) Left: Prefill Latency (to encode a full prompt); (b) Right: Decode Latency per token.
    }
    \label{latency}
\end{figure*}

In the prefill stage, the model processes many tokens at once. This creates large matrix multiplications that are compute-intensive. Prefill latency decreases with both frequency and local buffer size, and our initial experiments showed that performance does not improve with local buffer sizes greater than 256KB. This pattern shows that prefill is compute-bound. While the number of cycles in Prefill changes little with $f$, an important clarification here is that the actual latency (Delay) for Prefill (being \emph{compute-bound}) is strongly dependent on frequency (see Figure~\ref{latency}a, explained below). This happens because the frequency scaling from 200 to 1400 (7$\times$) outpaces the decrease in the cycle count ($6.0\times10^8~to~3.0\times10^8$, only 2$\times$). In other words, while the total number of cycles decreases with f, the actual latency increases.

In the decode stage, generation happens one token at a time. This drastically reduces arithmetic intensity, since compute scales down, but memory accesses for reading weights does not. Each new token still requires running the entire transformer stack which includes Attention, i.e. Q, K, V projections, KV cache lookups, softmax, etc., and MLP projections. The model's weights are often tens to hundreds of GB, and even though only one token is processed, the full weight matrices must be streamed from memory. Compute per token is relatively small, but memory reads per token are massive and so memory bandwidth dominates. The number of cycles in decode (Figure~\ref{fig:decode_cycle_count}) increase with $f$ for $f > $ 400MHz, while decode latency is unaffected by frequency for $f > $ 400MHz. When $f$ is greater than 400MHz, decode takes more cycles to complete yet latency stays constant (see Figure~\ref{latency}b, explained below). It is clear that decode is compute bound for $f <= $ 400MHz and memory bound for $f >= $ 600MHz.
%Consequently, there seems to exist an ideal frequency for this workload and hardware between 400MHz and 600MHz that can achieve the best possible latency while minimizing wasted cycles. 

% Grid showing prefill vs decode analyses (page 2)

\begin{figure*}[ht]
  \centering
    \begin{subfigure}{0.48\textwidth}
    \centering
    \includegraphics[width=\linewidth]{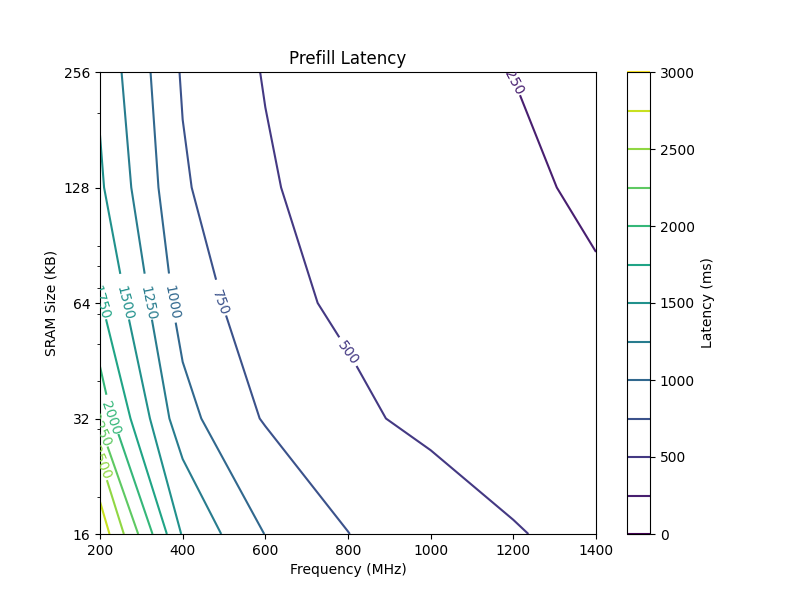}
    %\caption{Prefill Latency Isoplot}
    %\label{prefill_latency_isoplot}
  \end{subfigure}
      \hfill
  \begin{subfigure}{0.48\textwidth}
    \centering
    \includegraphics[width=\linewidth]{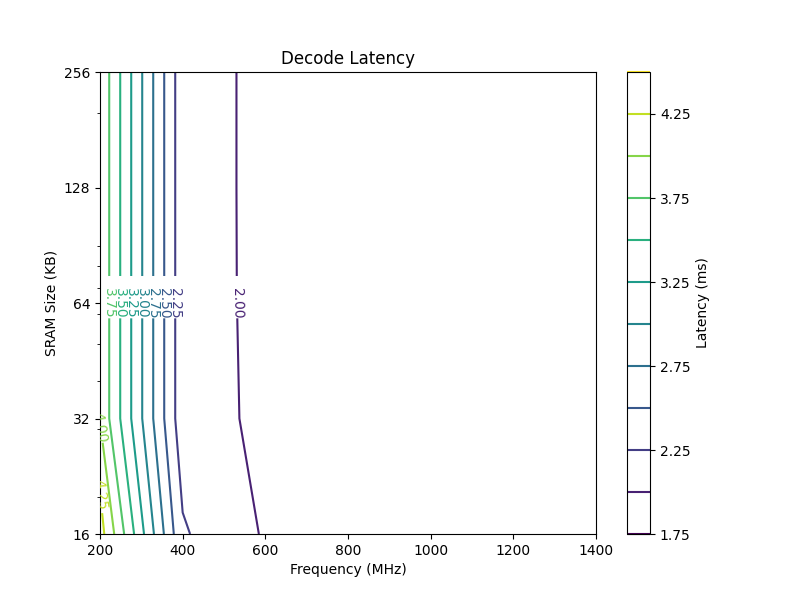}
   % \caption{Decode Latency per Token Isoplot}
    %\label{decode_latency_isoplot}
  \end{subfigure}
  
  \caption{Isoplot Analysis: (a) Left: Prefill Latency (to encode a full prompt); (b) Right: Decode Latency per token}
\label{isoplot-latency}
\end{figure*}

\subsection{Latency}

%Figures \ref{prefill_latency}, \ref{decode_latency}, 
%Figures \ref{latency}a, \ref{latency}b, 
%\ref{dynamic_energy}a, 
%\ref{dynamic_energy}b, \ref{static_energy}a, \ref{static_energy}b, \ref{total_energy}, and %\ref{total_energy} 

Figures \ref{latency}, \ref{dynamic_power}, \ref{dynamic_energy}, \ref{static_energy}, and \ref{total_energy}, show latency, power, and energy metrics for local buffer size, $S$, of 16KB, 32KB, 64KB, 128KB, 256KB, 512KB, and 1024KB. For each SRAM size, there are seven bars that represent the seven frequencies, $f$, 200MHz--1400MHz (in steps of 200MHz).

% Latency % 
Figures \ref{latency}a and \ref{latency}b show the latency for each frequency and local buffer size, for prefill and decode respectively. For the prefill stage, increasing the local buffer size from 16KB to 32KB shows a significant reduction in latency. However, further increases in buffer size (e.g., from 32KB to 64KB, 128KB, 256KB) result in progressively smaller reductions in latency, suggesting diminishing returns after 32KB. The latency values for 32KB, 64KB, 128KB, and 256KB are similar, especially at higher frequencies. In decode,  increasing the local buffer size from 16KB to 32KB reduces decode latency at the lowest compute frequencies, but decode is otherwise not affected by increasing buffer size, $S$, or frequency, $f$.
%$S$ or $f$.

% Describe decode latency graph %

%Therefore, decode is memory bound and not sensitive to $f$ for $f$ $>$ 400MHz. 
When frequency is too low, decode becomes compute bound instead of memory bound since computation latency surpasses memory latency. SRAM size does not affect decode latency, except for with 16 KB SRAM where there is an increase in latency for $f$ $<=$ 400MHz. This shows that SRAM size can affect decode latency if decode becomes compute-bound due to low compute frequency.

Figure~\ref{isoplot-latency} shows a different view of the latency results, a latency isoplot, demonstrating the combinations of $S$ and $f$ that result in approximately the same latency. Figure~\ref{isoplot-latency}a shows the isoplot for prefill and \ref{isoplot-latency}b for decode, respectively. 
The isoplots show that prefill latency is dependent on both SRAM size and frequency. The lowest latency regions (dark blue/purple) are found at the highest frequencies (right side of the plot) and larger SRAM sizes (top part of the plot). Specifically, combinations of high frequency (e.g., 1200-1400 MHz) and SRAM sizes 32KB and above tend to show the best latency. The contour lines are increasingly compressed for SRAM sizes between 16KB and 32KB, especially at lower frequencies, indicating a rapid decrease in latency in these areas. However, for SRAM sizes above 32KB, the contour lines become much more spread out horizontally, particularly towards the higher frequencies. %This signifies that increasing SRAM size beyond 32KB has a diminishing impact on latency reduction, even at higher frequencies.
Figure~\ref{isoplot-latency}b shows that decode latency is largely independent of SRAM size and can only be decreased by increasing frequency.

\subsection{Compute Percent of Latency}
%%%%%% Compute percent of latency %%%%%%
\begin{figure}[ht]
    \centering
    \includegraphics[width=\linewidth]{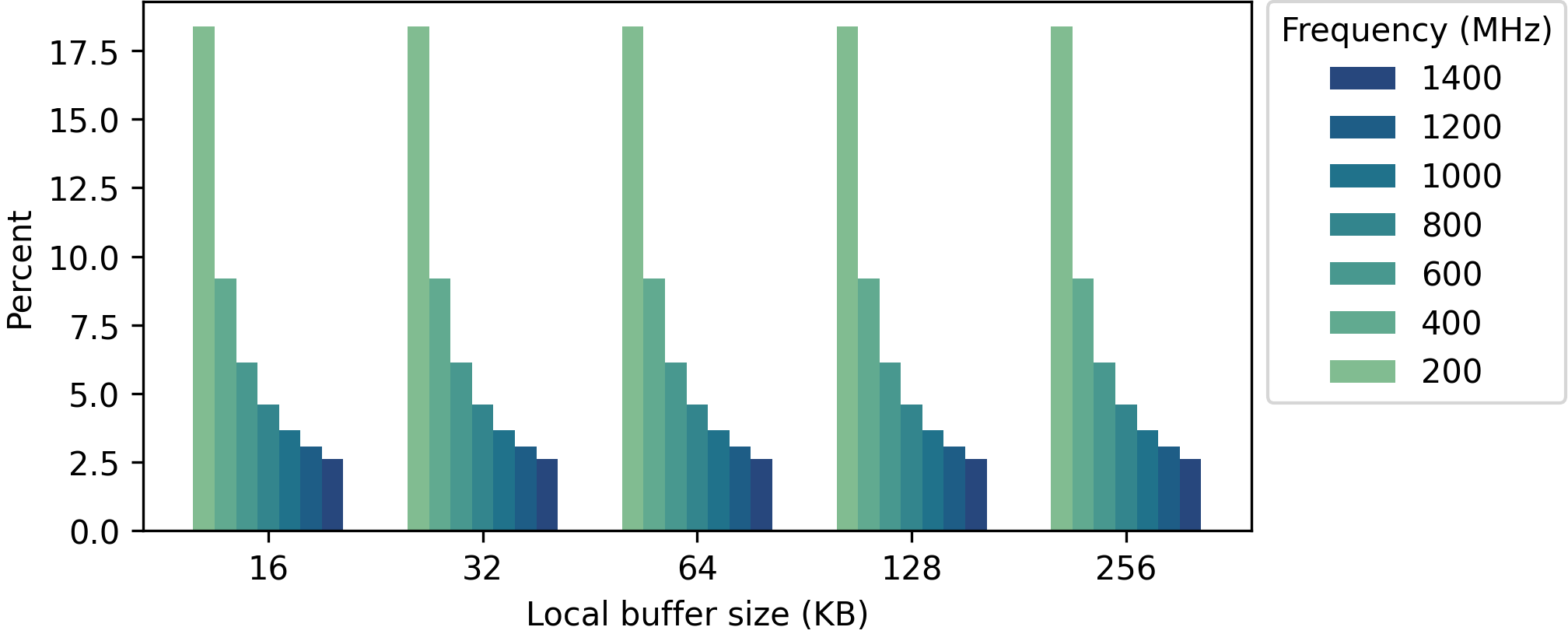}
    \caption{Compute Percent of Latency (decode)}
    \label{compute_percent}
\end{figure}

Figure \ref{compute_percent} shows how much of decode latency results from compute, where the remaining latency is caused by waiting on memory. The percent of latency attributable to compute is necessary to correctly calculate how much energy is used for computation. We modified the LLMCompass code to output the percentage of latency from compute.  In the prefill phase, nearly all of latency can be attributed to compute but in decode less than 20\% of latency can be attributed to compute because decode is heavily memory-bound. The compute percent of latency is greatly affected by frequency, but unaffected by SRAM size.

% Dynamic Power

\subsection{Dynamic Power}
\begin{figure*}[ht]
  \centering  
    \begin{subfigure}{0.48\textwidth}
    \centering
    \includegraphics[width=\linewidth]{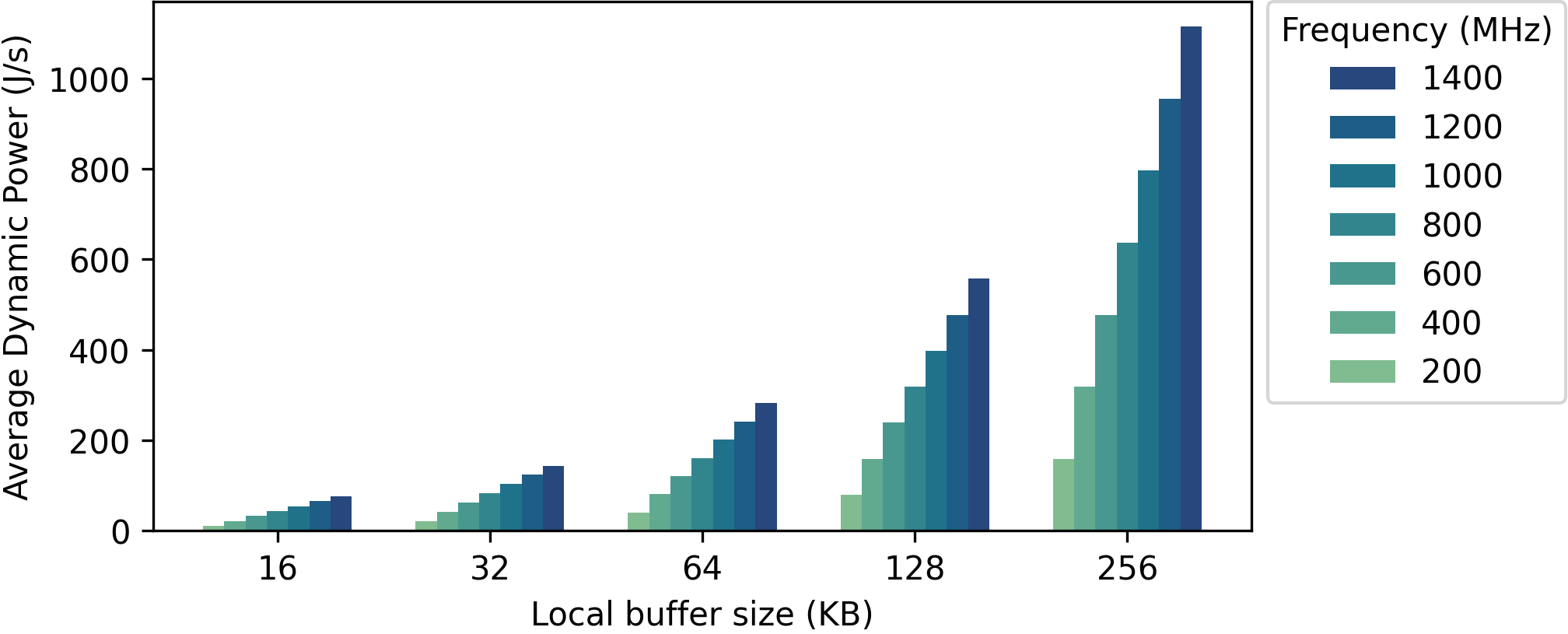}
  \end{subfigure}
  \hfill
  \begin{subfigure}{0.48\textwidth}
    \centering
    \includegraphics[width=\linewidth]{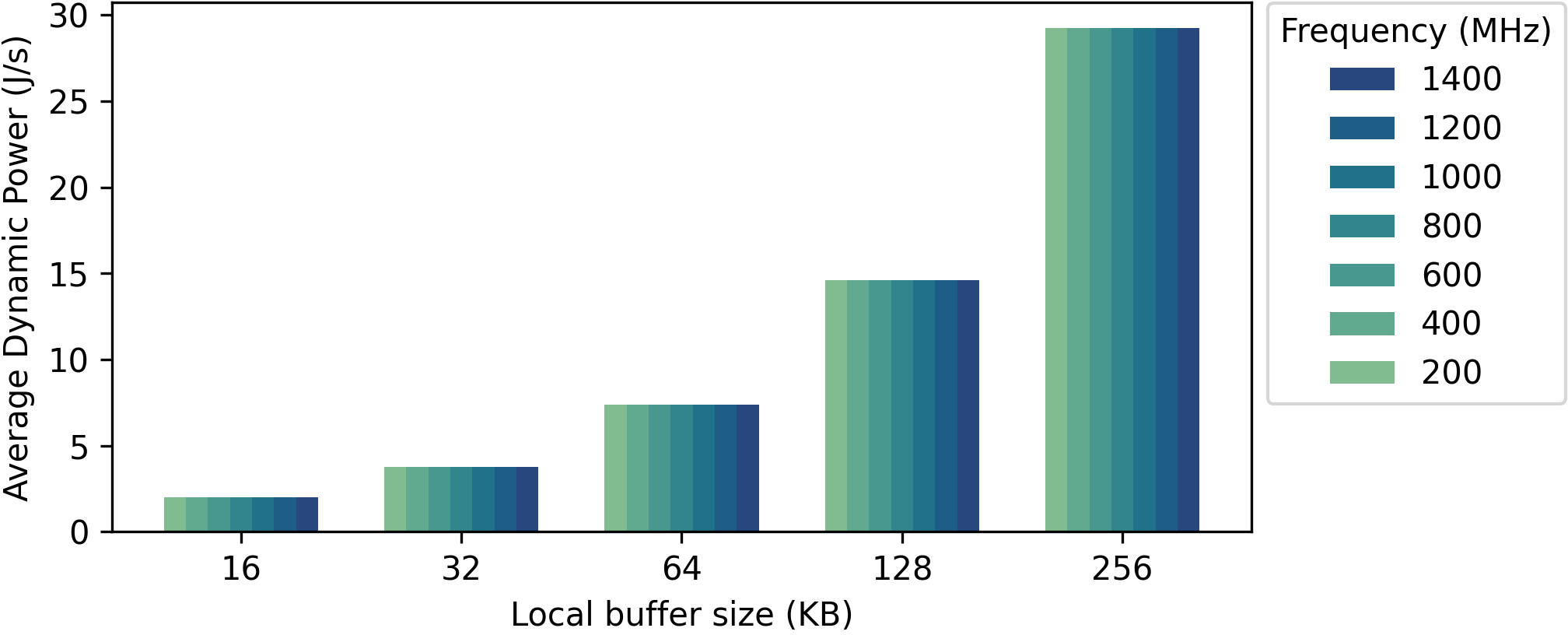}
  \end{subfigure}
      \caption{Dynamic power: (a) Left: Prefill (to encode a full prompt); (b) Right: Decode per token.}
      \label{dynamic_power}
  \end{figure*}

Figure \ref{dynamic_power}a shows that the dynamic power for prefill tends to increase linearly with frequency. Larger local buffer sizes generally show higher dynamic power due to increased activity. The power values are typically higher than decode due to the larger amount of data processing.
%\tv{for power the unit is Watt, for energy it is Joule}
Figure~\ref{dynamic_power}b shows that the dynamic power for decode increases with $S$ but is unaffected by frequency. %TODO explain why 

% Dynamic Energy  %
\subsection{Dynamic Energy}
 \begin{figure*}[ht]
  \centering 
\begin{subfigure}{0.48\textwidth}
    \centering
    \includegraphics[width=\linewidth]{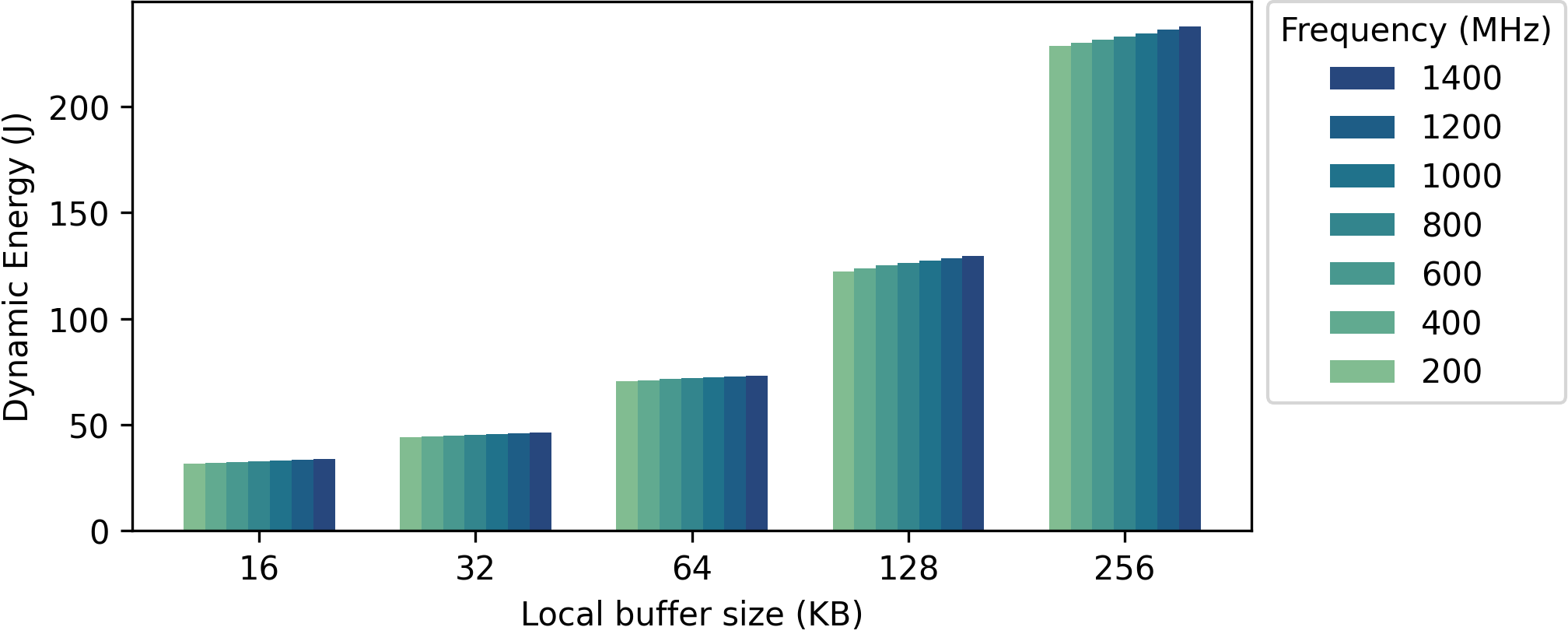}
  \end{subfigure}
  \hfill
  \begin{subfigure}{0.48\textwidth}
    \centering
    \includegraphics[width=\linewidth]{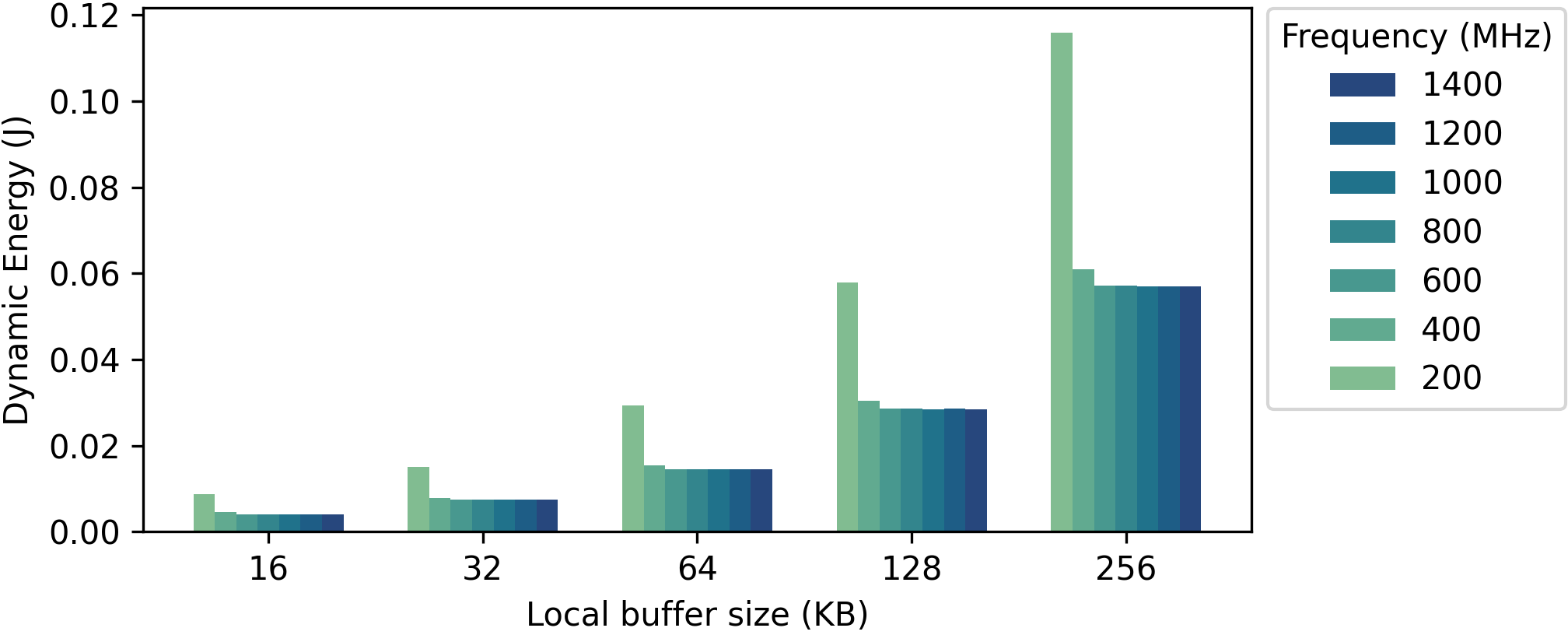}
  \end{subfigure}
  \caption{Dynamic energy: (a) Left: Prefill (to encode a full prompt); (b) Right: Decode per token.}
  \label{dynamic_energy}
\end{figure*}

Figure \ref{dynamic_energy}a shows dynamic energy consumption during prefill, and \ref{dynamic_energy}b during decode. Dynamic energy is primarily dependent on SRAM size, but frequency has a small effect on prefill while decode is unaffected by increasing frequency past 600MHz. Dynamic power expenditure increases for larger SRAMs because of how SRAM circuits scale electrically.
%: in an SRAM array, each row has a wordline, and each column has bitlines. 
As SRAM size increases, its subarrays increase both in size (wordlines and bitlines become longer) but also in number (requiring longer interconnects). Charging and discharging these capacitive lines is the main reason for SRAM dynamic power dissipation~\cite{IJECCE297}, and longer interconnects use more energy per access.

% Static Energy %
\subsection{Static Energy}
\begin{figure*}[ht]
  \centering
    \begin{subfigure}{0.48\textwidth}
    \centering
    \includegraphics[width=\linewidth]{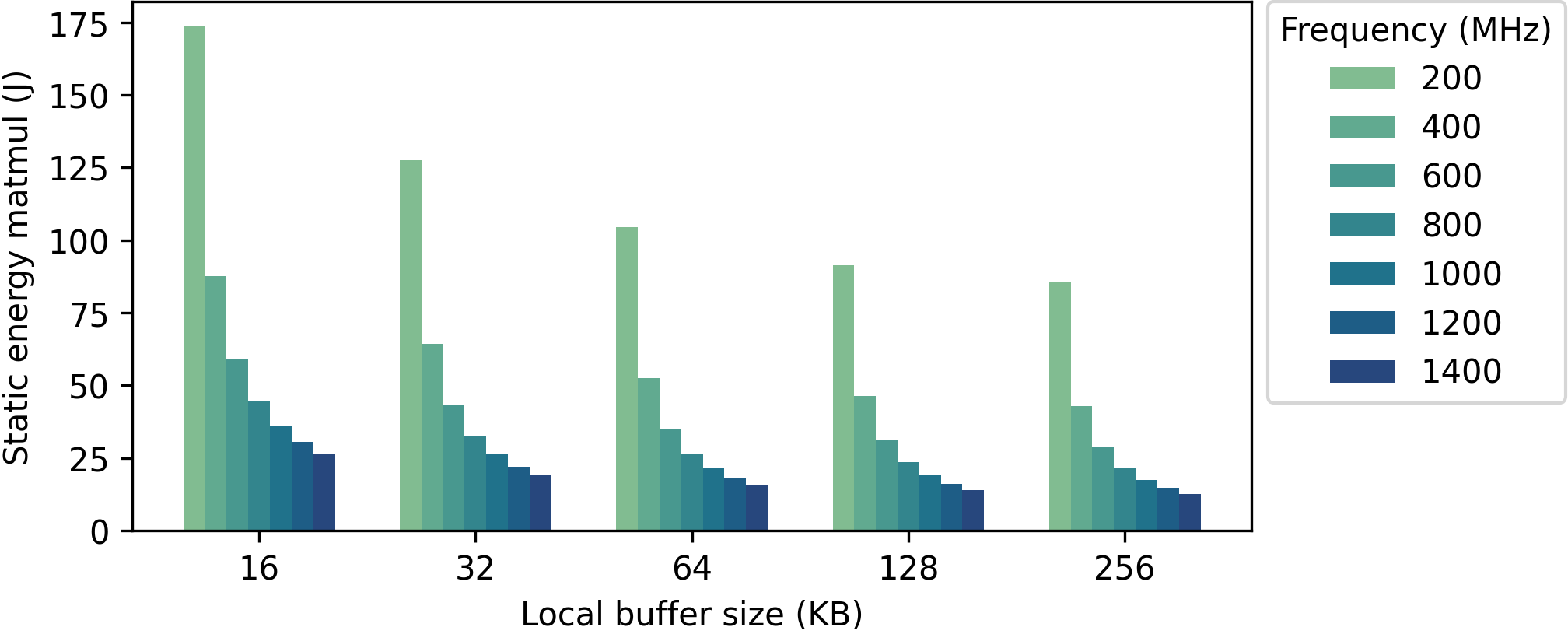}
  \end{subfigure}
  \hfill
  \begin{subfigure}{0.48\textwidth}
    \centering
    \includegraphics[width=\linewidth]{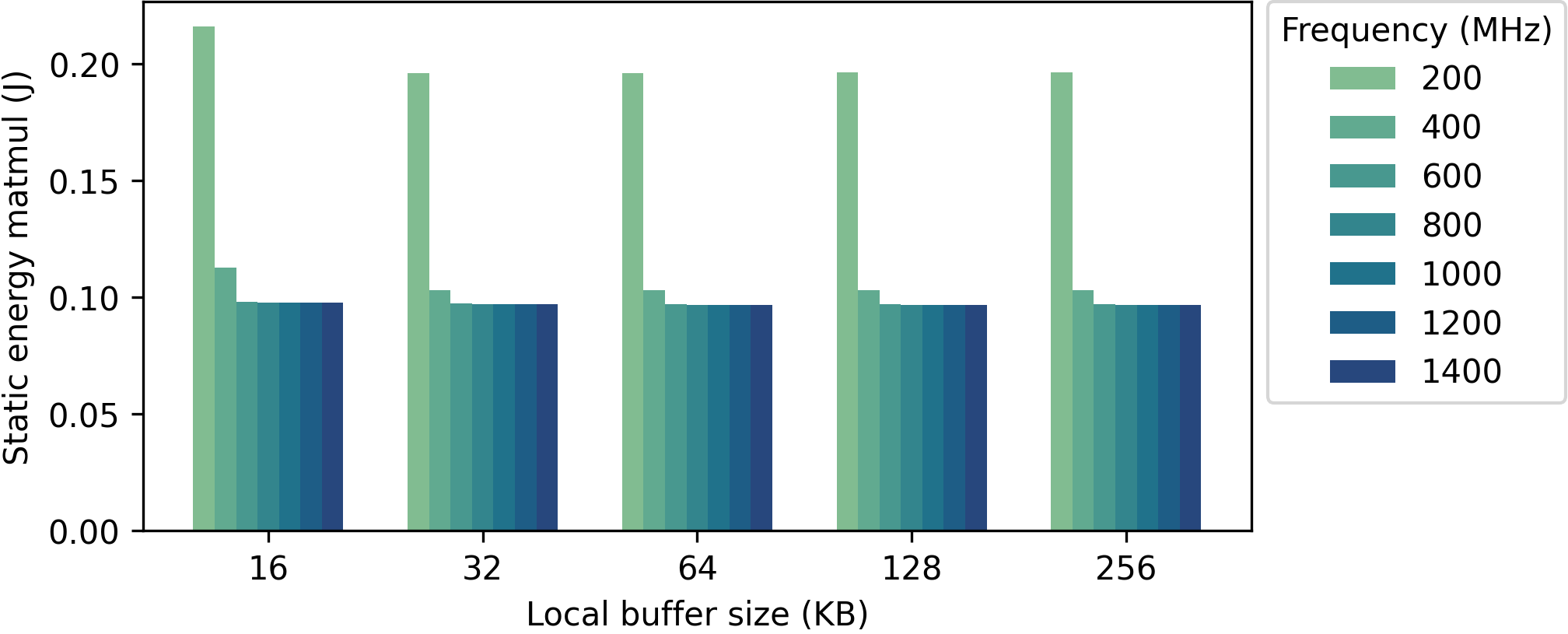}
  \end{subfigure}
  \caption{Static Energy: (a) Left: Prefill (to encode a full prompt); (b) Right: Decode per token.}
  \label{static_energy}
\end{figure*}

Figures \ref{static_energy}a and \ref{static_energy}b show static energy for each $S$ and each $f$. We calculate static energy by multiplying latency by leakage power of the local buffers, global buffer, and systolic arrays. As expected in line with the latency results, in prefill, static energy increases significantly with $S$, and decreases as frequency increases. This is due to the decreased amount of time spent in the prefill stage due to latency improvements from increased frequency. In decode, static energy is only affected by SRAM size, except when frequency is too low such that decode latency becomes compute bound, causing increased latency. %In other words, prefill is generally compute bound while decode is generally memory bound.

% Total Energy %
\subsection{Putting it all together: Total Energy}
\begin{figure*}[ht]
  \centering  
    \begin{subfigure}{0.48\textwidth}
    \centering
    \includegraphics[width=\linewidth]{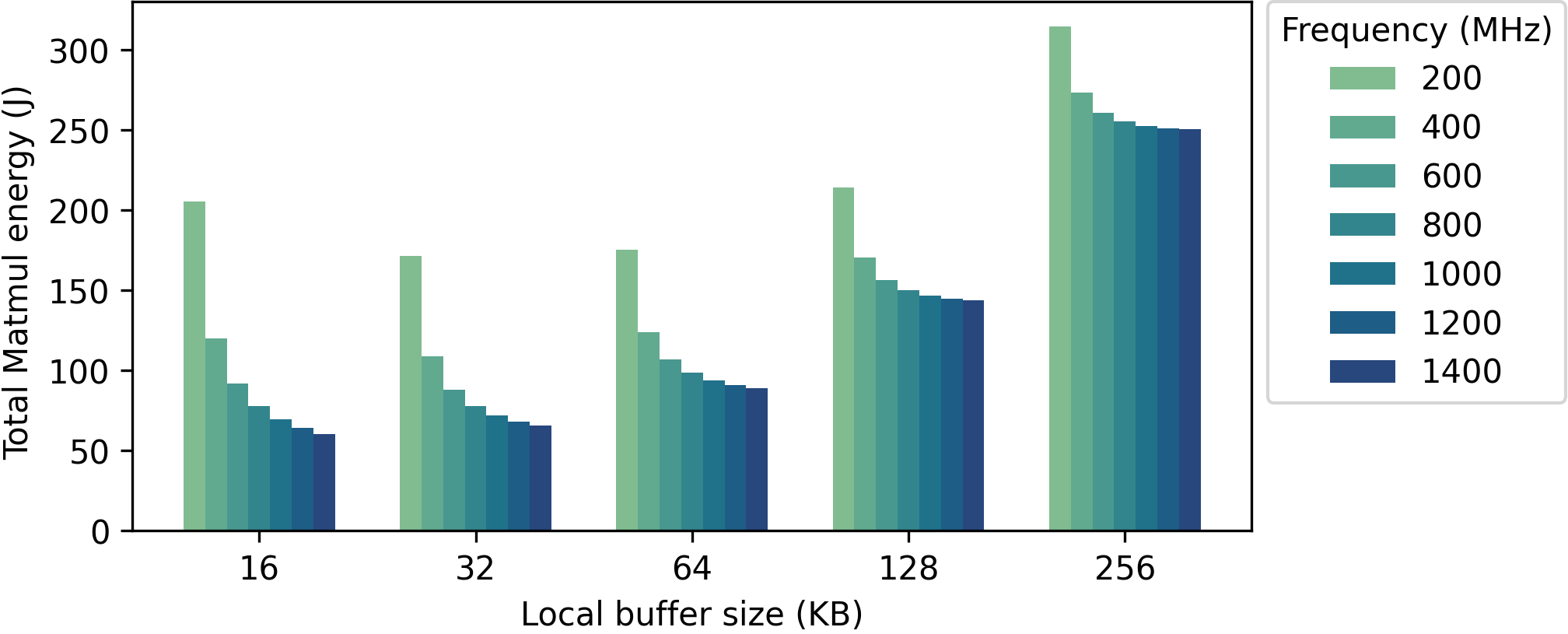}
  \end{subfigure}
  \hfill
  \begin{subfigure}{0.48\textwidth}
    \centering
    \includegraphics[width=\linewidth]{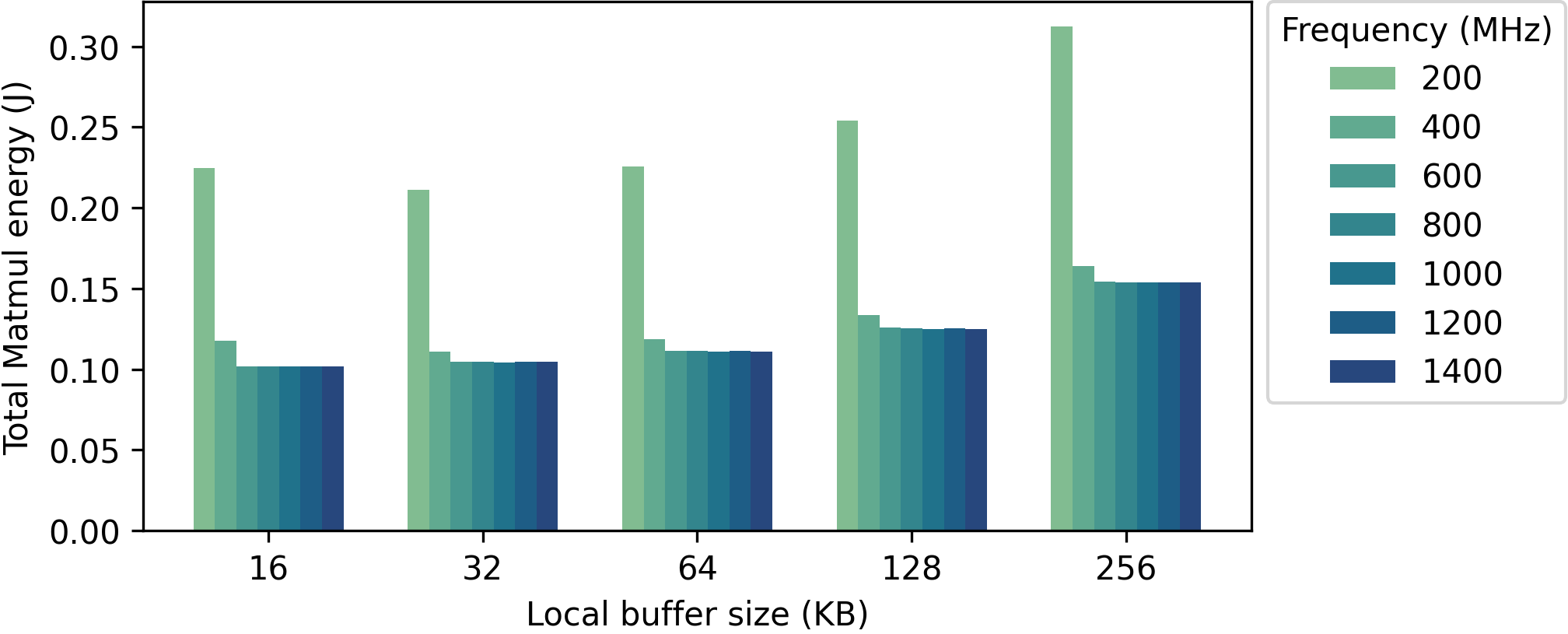}
  \end{subfigure}
  \caption{Total Energy: (a) Left: Prefill (to encode a full prompt); (b) Right: Decode per token.}
  \label{total_energy}
\end{figure*}

\begin{figure*}[ht]
  \begin{subfigure}{0.48\textwidth}
    \centering
    \includegraphics[width=\linewidth]{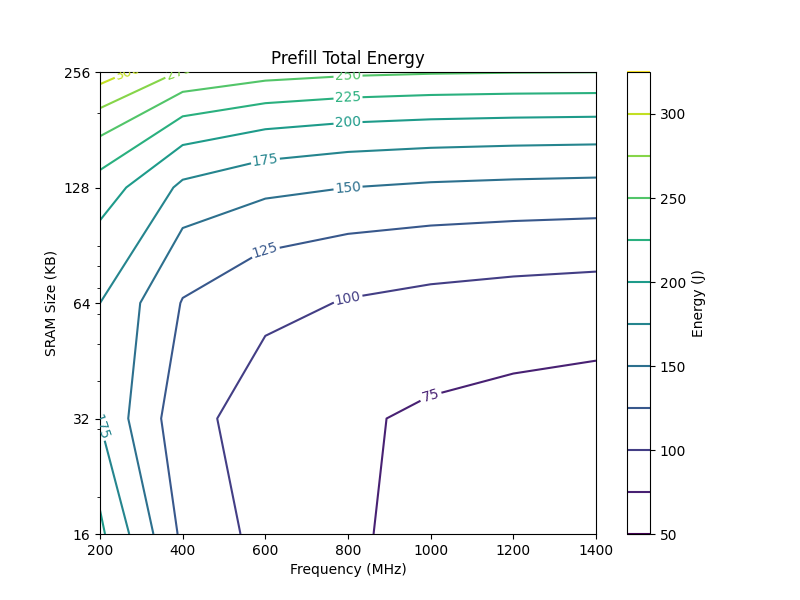}
    %\caption{Prefill Total Energy Isoplot}
    %\label{prefill_energy_isoplot}
  \end{subfigure}
      \hfill
  \begin{subfigure}{0.48\textwidth}
    \centering
    \includegraphics[width=\linewidth]{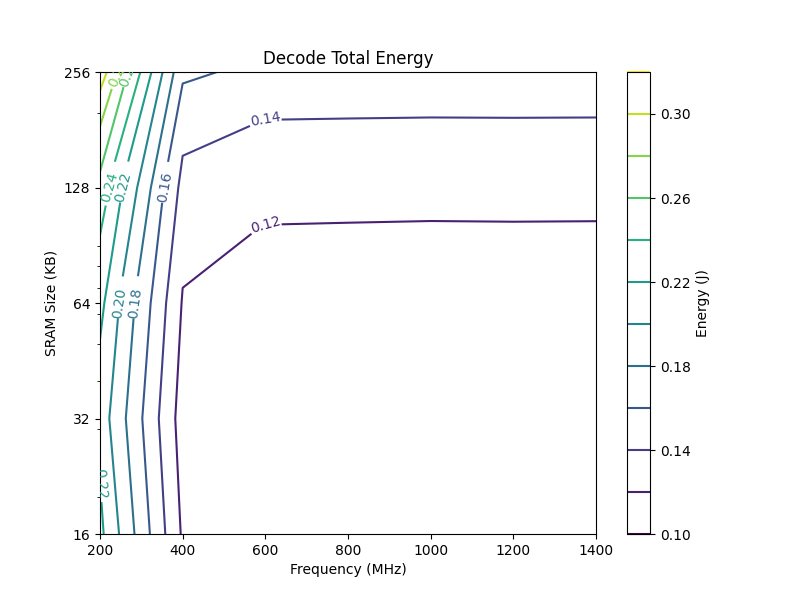}
    %\caption{Decode Total Energy per Token Isoplot}
    %\label{decode_energy_isoplot}
  \end{subfigure}
    
  \caption{Isoplot Analysis: (a) Left: Prefill Total Energy (to encode a full prompt); (b) Right: Decode Total Energy per token.}
\label{isoplot-energy}
\end{figure*}

Figures \ref{total_energy}a and \ref{total_energy}b show total energy for each SRAM size and $f$. Total energy is calculated by adding static energy to dynamic energy. The results show that SRAM size affects total energy use during both prefill and decode, and that frequency has a slight effect on prefill energy use and no effect on decode energy use, except for at the compute-bound lowest frequencies. We note that prefill and decode both achieve lowest total energy for all frequencies with a 32KB local SRAM.

Figures \ref{isoplot-energy}a and \ref{isoplot-energy}b show the isoplots for total energy for prefill and decode, respectively. Prefill total energy depends on changes in both $S$ and $f$ showing that one can choose more than one different combination of $S$ and $f$ to achieve a target energy consumption for a prefill task. 
However, running too slow is expensive and large SRAM is an energy tax in prefill.
In contrast, decode energy per token drops sharply as frequency increases up to a knee around 400 MHz, then largely plateaus with frequency (above 600 MHz) while increasing mainly with larger SRAM. In other words, above the knee, $S$, not $f$, dominates energy. 

\iffalse
Prefill total energy (left)
What the contours look like: strong “knee” at low frequency, then mostly horizontal bands as SRAM grows.
Conclusions
Running too slow is expensive (race-to-halt effect).
At ~200–350 MHz, you cross multiple energy contours quickly: lower frequency stretches runtime, so total energy rises (typically leakage + “keep-everything-on-longer” overhead dominates).
Large SRAM is an energy tax in prefill.
Past moderate SRAM sizes, contours become largely horizontal: energy increases substantially as SRAM increases, and frequency doesn’t rescue you much. That’s consistent with SRAM’s static/leakage and access energy becoming a big part of prefill’s total energy.
Best prefill-energy region is “moderate/high f + small/moderate SRAM.”
The lowest-energy area sits toward the bottom-right-ish (higher frequency, smaller SRAM). Increasing SRAM beyond what you need for reuse gives steadily worse energy with limited benefit.

\fi

\subsection{Energy Delay Product}
\begin{figure*}[ht]
  \centering

  \begin{subfigure}{0.48\textwidth}
    \centering
    \includegraphics[width=\linewidth]{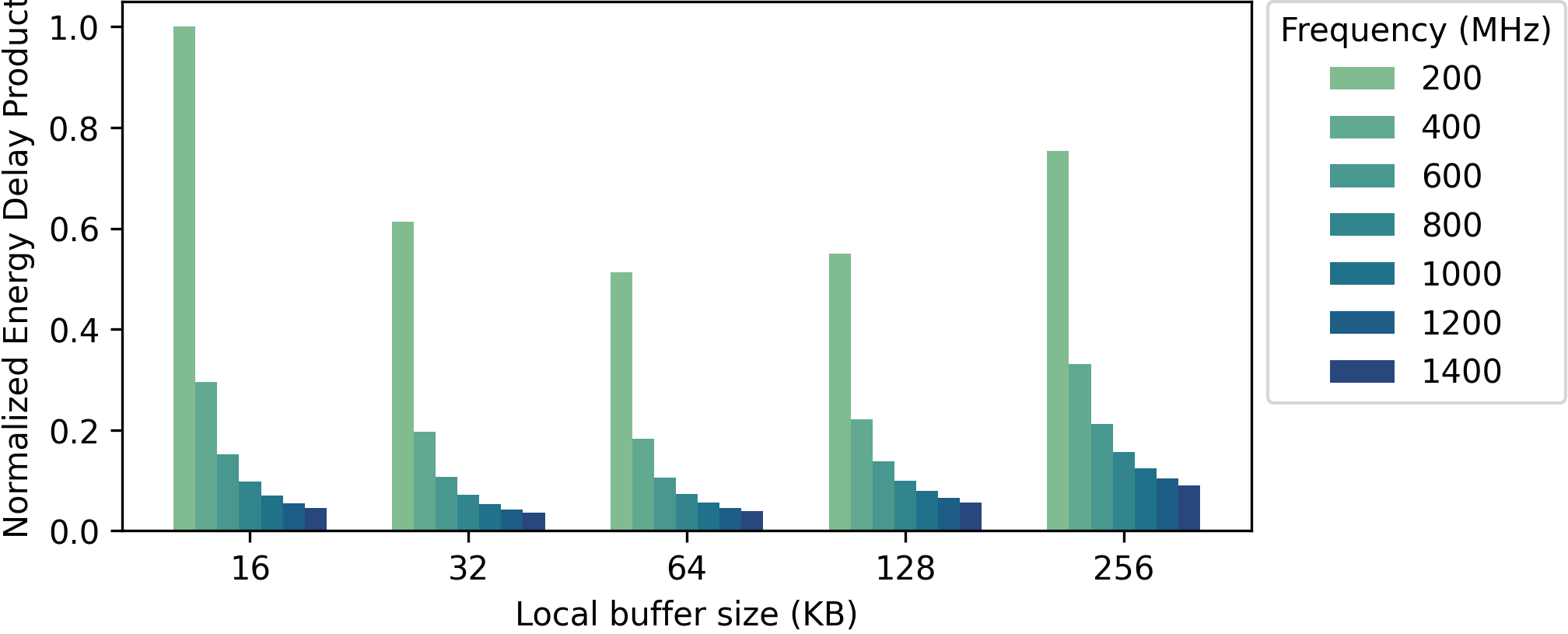}
    %\caption{Prefill Energy Delay Product}
  \end{subfigure}
      \hfill
  \begin{subfigure}{0.48\textwidth}
    \centering
    \includegraphics[width=\linewidth]{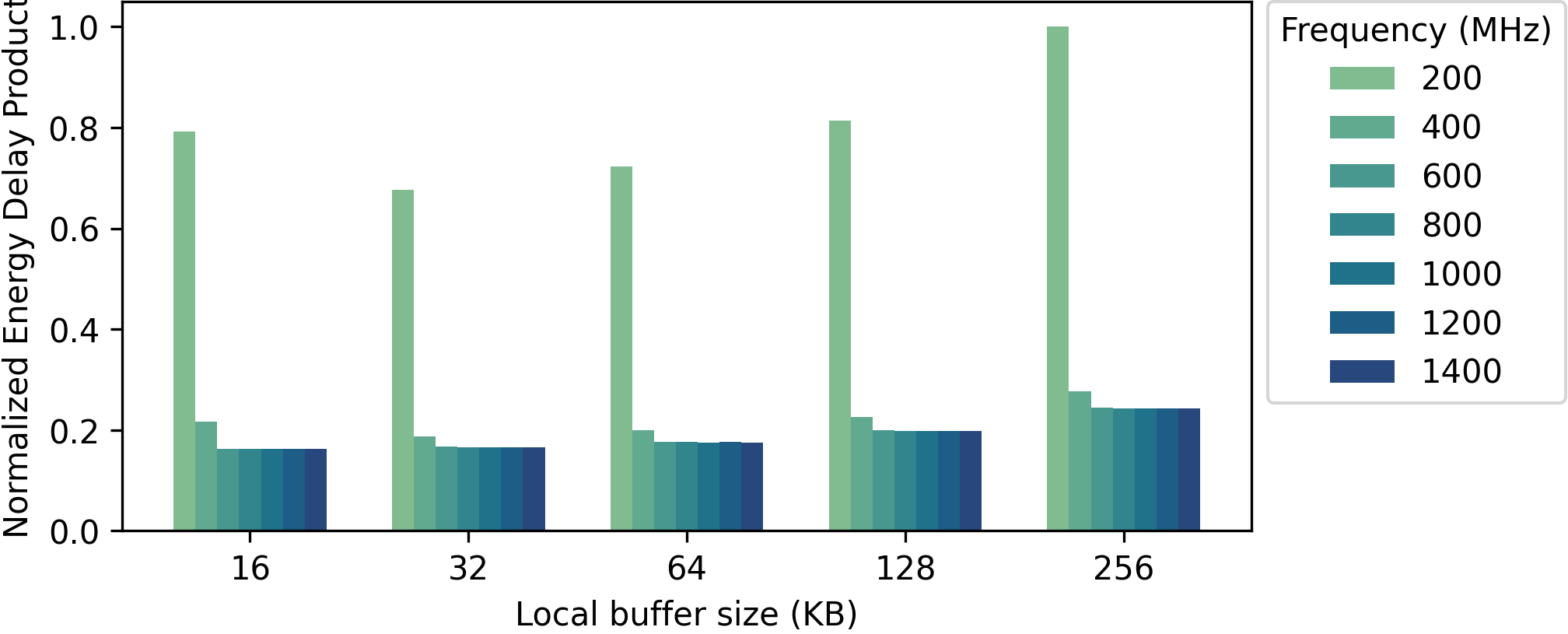}
    %\caption{Decode Energy Delay Product}
  \end{subfigure}
\caption{Energy Delay Product: (a) Left: Prefill (to encode a full prompt); (b) Right: Decode per token.}
\label{edp}
\end{figure*}

Figure \ref{edp}a and \ref{edp}b show the energy-delay product for prefill and decode respectively. The energy–delay product is a composite efficiency metric that captures the trade-off between energy per operation and speed of operation, and is normalized. A low EDP indicates that a frequency and SRAM size combination is both energy-efficient and fast, while a high EDP indicates that the configuration is either too energy-hungry, too slow, or both. In prefill, EDP decreases as $S$ increases from 16KB to 64KB. However, EDP increases for $S >$ 64KB. Prefill EDP decreases with increased frequency for all $S$, because of decreased latency.
In decode, we see a drop in EDP for lower compute frequencies as $S$ increases from 16KB to 32KB, and an increase in EDP for $S >$ 32KB. Decode EDP is not sensitive to increasing frequency for $f >$ 400MHz, because compute is such a small percent of latency.

\subsection{Roofline}
% Roofline %
\begin{figure*}[ht]
  \centering
  \begin{subfigure}{0.48\textwidth}
    \centering
    \includegraphics[width=\linewidth]{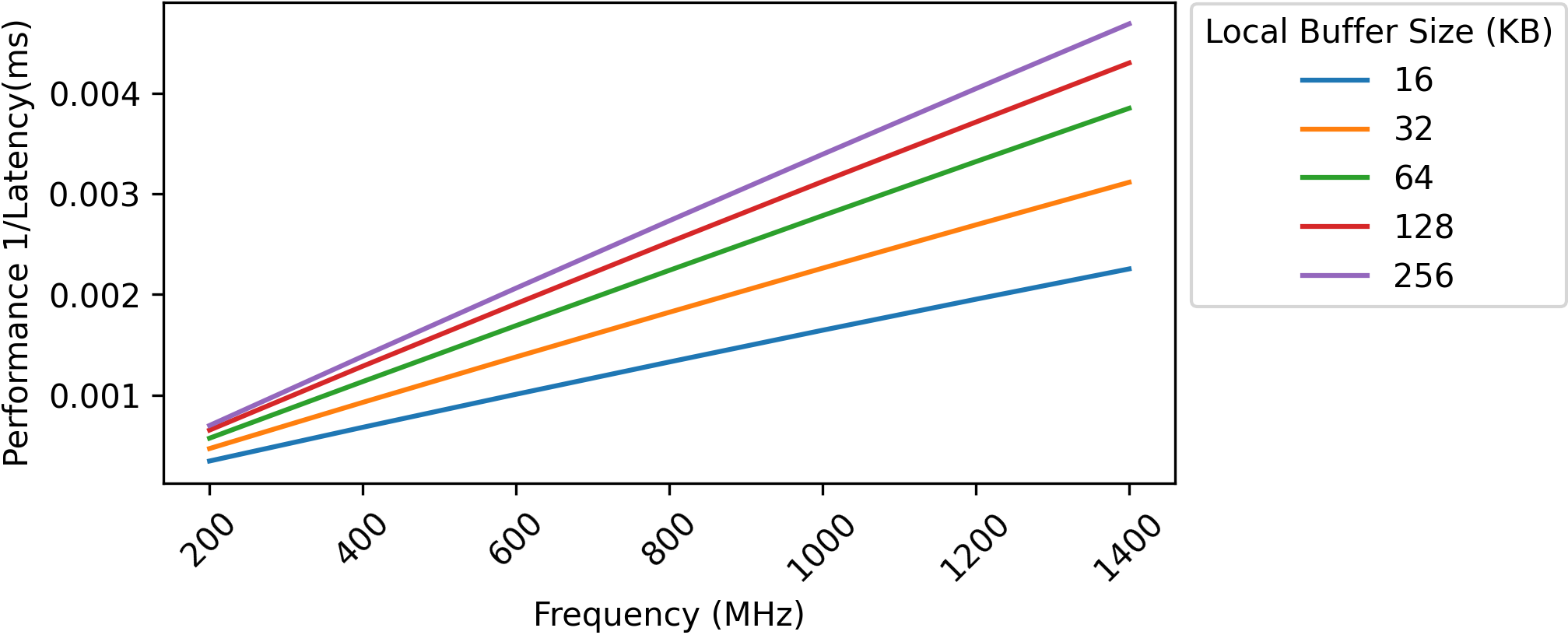}
    %\caption{Roofline Prefill}
  \end{subfigure}
      \hfill
  \begin{subfigure}{0.48\textwidth}
    \centering
    \includegraphics[width=\linewidth]{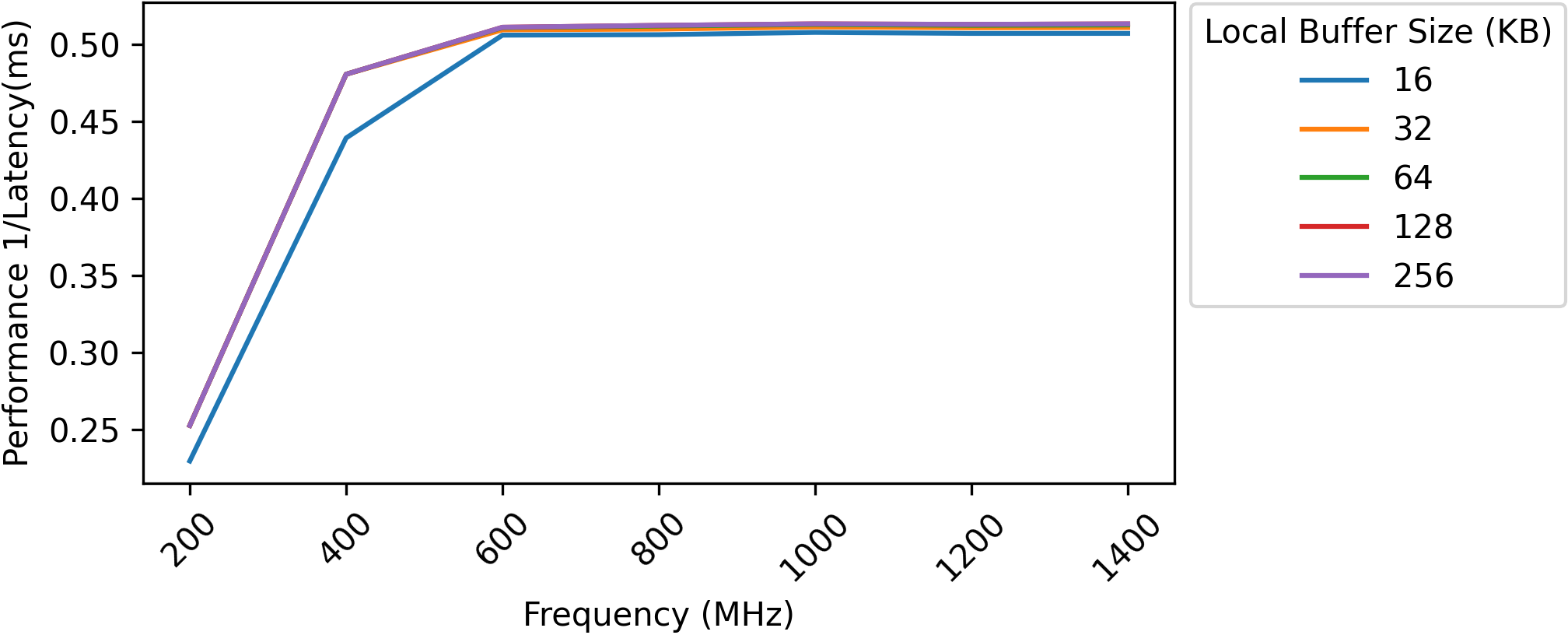}
    %\caption{Roofline Decode}
  \end{subfigure}
\caption{Roofline Analyses: (a) Left: Prefill (to encode a full prompt); (b) Right: Decode per token.}
\label{roofline}
\end{figure*}

Figures \ref{roofline}a and \ref{roofline}b show roofline graphs for prefill and decode. Performance is given as the reciprocal of latency for processing the entire prompt for prefill and per token for decode. The roofline model helps determine whether a program is compute-bound or memory-bound. A memory-bound program's performance is limited by memory bandwidth, making low operational intensity preferred to conserve resources. A compute-bound program's performance is limited by compute resources, making high operational intensity preferred to minimize latency. Prefill's latency pattern is compute bound for all $S$ and $f$, and therefore we do not see a "roofline" capping performance. Decode's latency pattern can be explained with the roofline model: the latency increases at $S=$16KB, with increasing compute frequency compensating slightly for the smaller SRAM. SRAM sizes greater than 16KB do not improve performance, indicating a shift from compute-bound to memory-bound performance.

\begin{figure*}[ht]
  \centering
  \begin{subfigure}{0.32\textwidth}
    \centering
    \includegraphics[width=\linewidth]{images/roofline_decode.png}
    \caption{Roofline Model Baseline Memory Bandwidth}
    \label{roofline_baseline_memory_bandwidth}
  \end{subfigure}
      \hfill
  \begin{subfigure}{0.32\textwidth}
    \centering
    \includegraphics[width=\linewidth]{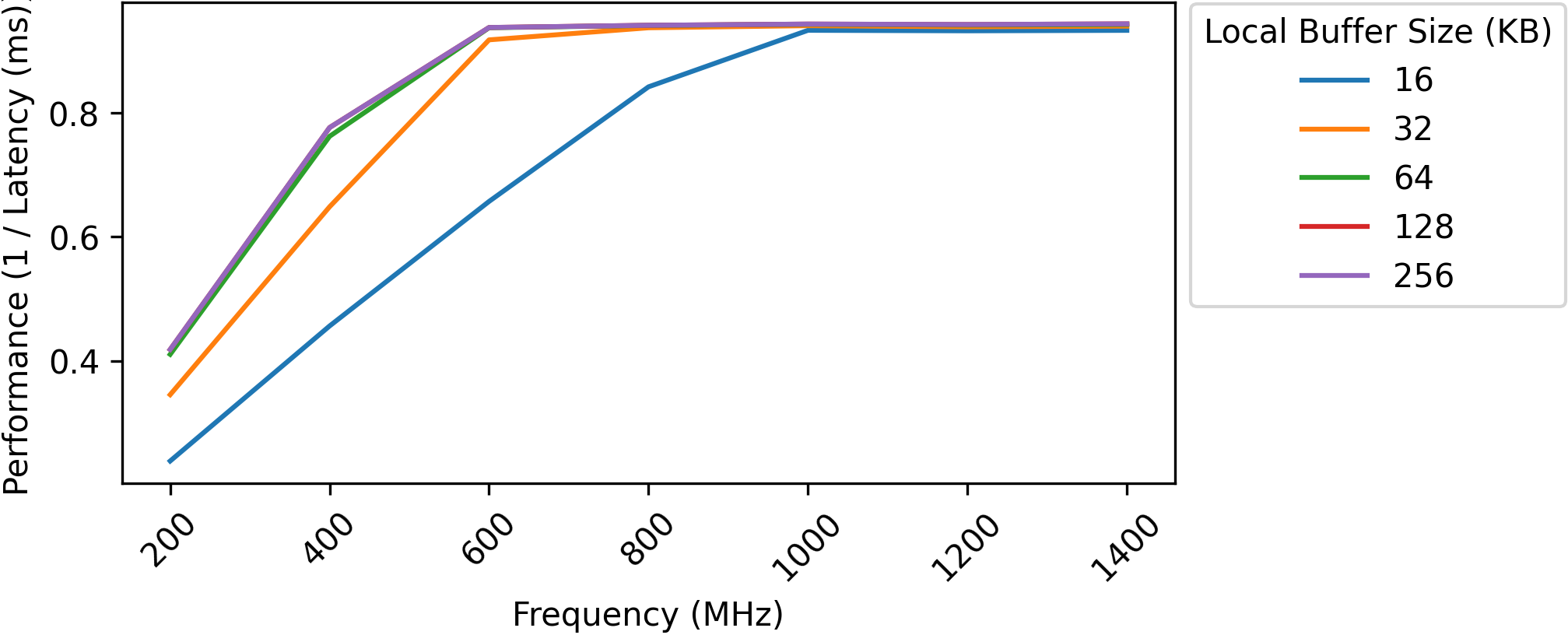}
    \caption{Roofline Model Double Memory Bandwidth}
    \label{roofline_double_memory_bandwidth}
  \end{subfigure}
      \hfill
  \begin{subfigure}{0.32\textwidth}
    \centering
    \includegraphics[width=\linewidth]{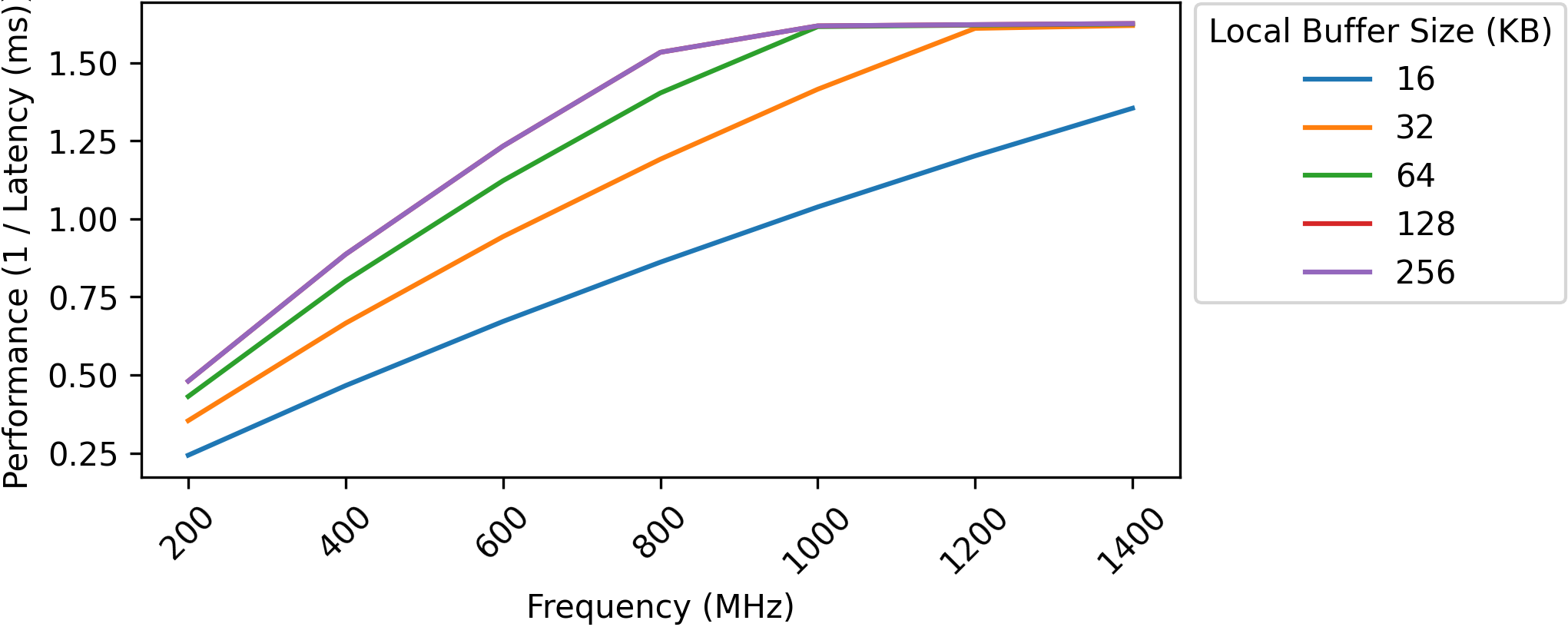}
    \caption{Roofline Model Quadruple Memory Bandwidth}
    \label{roofline_quadruple_memory_bandwidth}
  \end{subfigure}

  % Row 3
  \begin{subfigure}{0.32\textwidth}
    \centering
    \includegraphics[width=\linewidth]{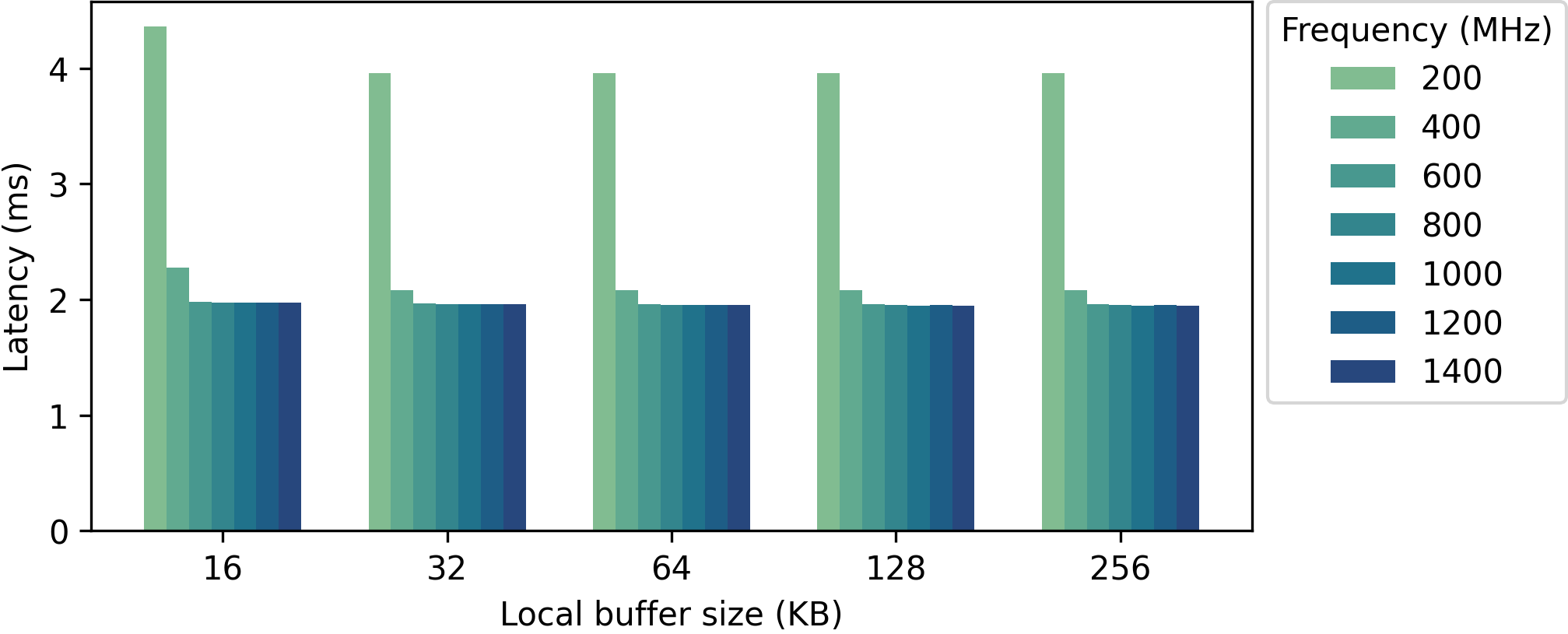}
    \caption{Decode Latency Baseline Memory Bandwidth}
    \label{decode_latency_baseline_memory_bandwidth}
  \end{subfigure}
  \hfill
  \begin{subfigure}{0.32\textwidth}
    \centering
    \includegraphics[width=\linewidth]{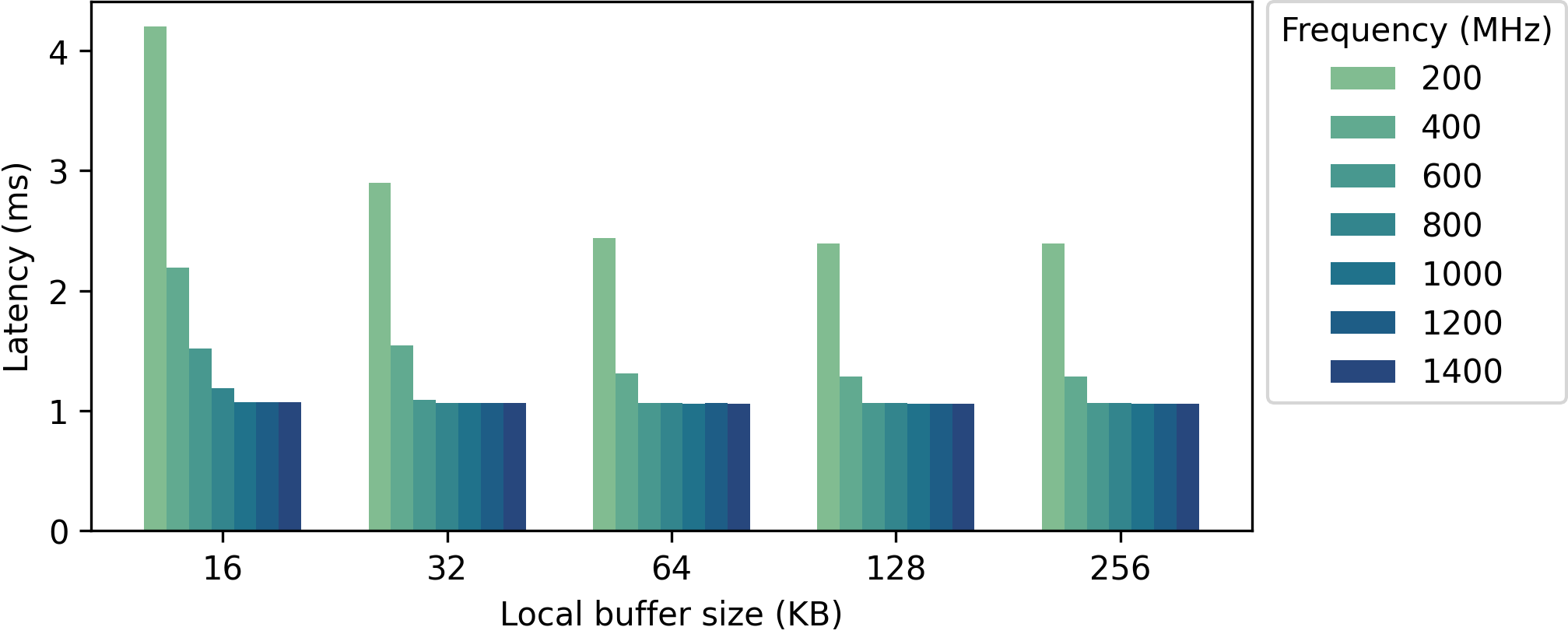}
    \caption{Decode Latency Double Memory Bandwidth}
    \label{latency_decode_double_memory_bandwidth}
  \end{subfigure}
  \hfill
    \begin{subfigure}{0.32\textwidth}
    \centering
    \includegraphics[width=\linewidth]{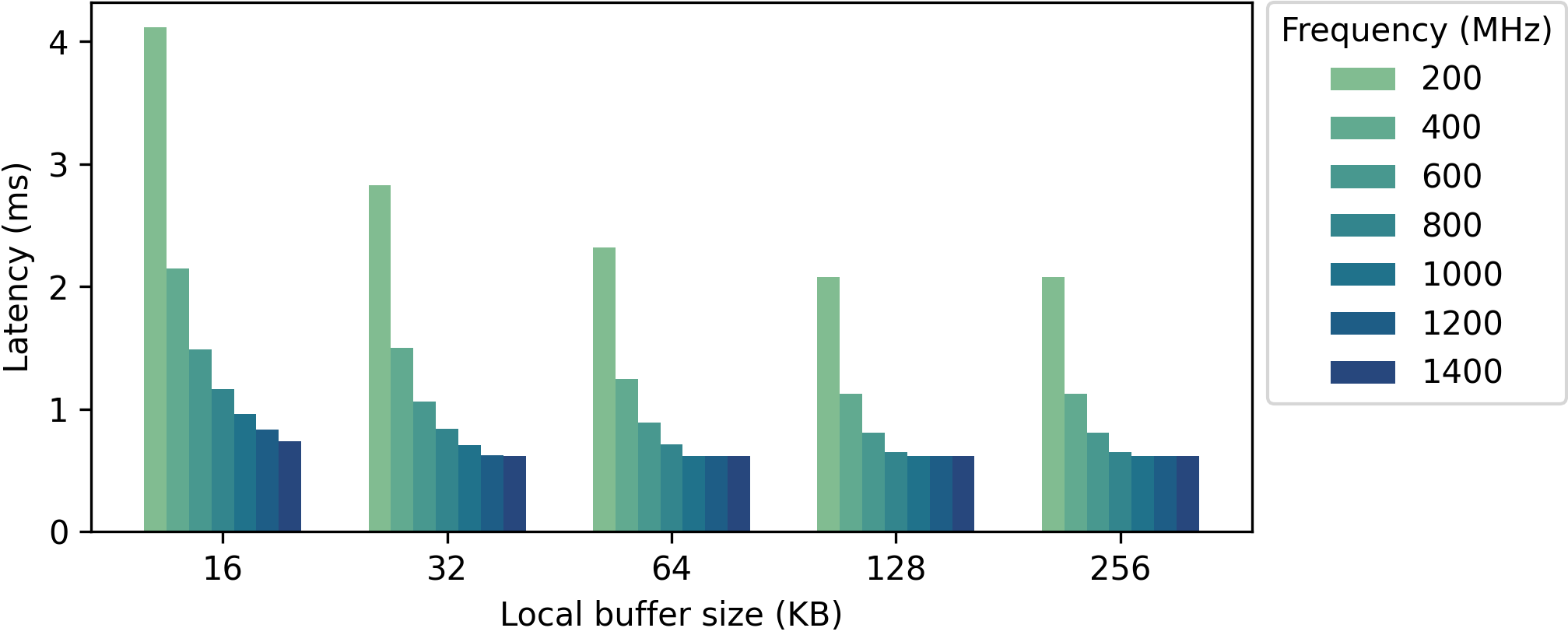}
    \caption{Decode Latency Quadruple Memory Bandwidth}
    \label{latency_decode_quadruple_memory_bandwidth}
  \end{subfigure}
  
  \begin{subfigure}{0.32\textwidth}
    \centering
    \includegraphics[width=\linewidth]{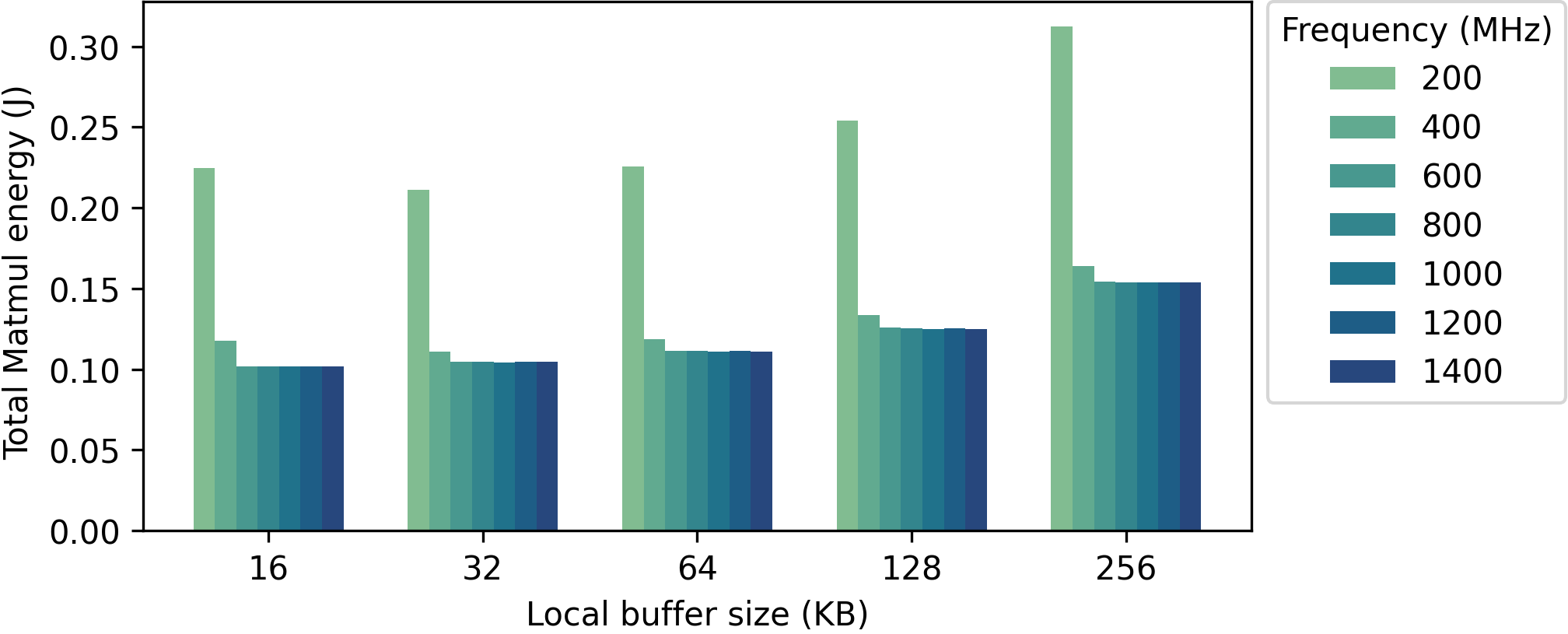}
    \caption{Decode Total Energy Baseline Memory Bandwidth}
    \label{decode_total_energy_baseline_memory_bandwidth}
  \end{subfigure}
    \hfill
  \begin{subfigure}{0.32\textwidth}
    \centering
    \includegraphics[width=\linewidth]{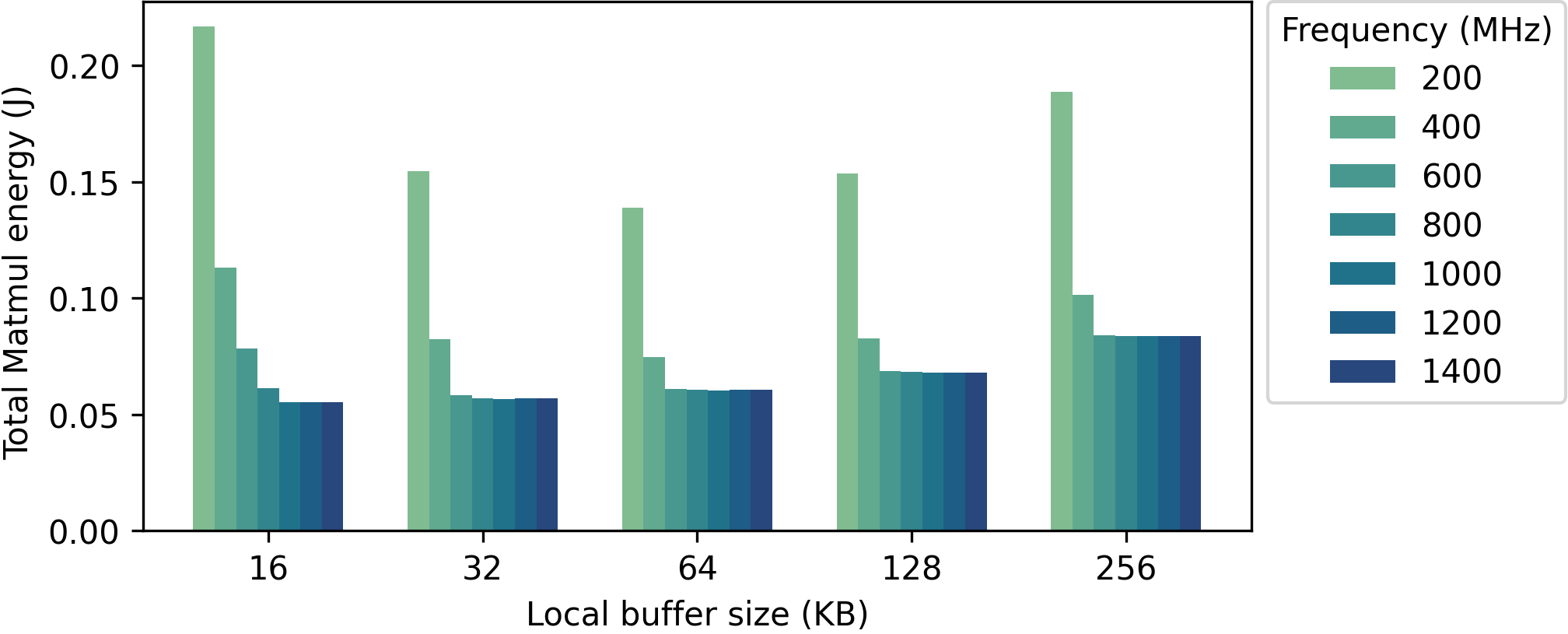}
    \caption{Decode Total Energy Double Memory Bandwidth}
    \label{decode_total_energy_double_memory_bandwidth}
  \end{subfigure}
      \hfill
        \begin{subfigure}{0.32\textwidth}
    \centering
    \includegraphics[width=\linewidth]{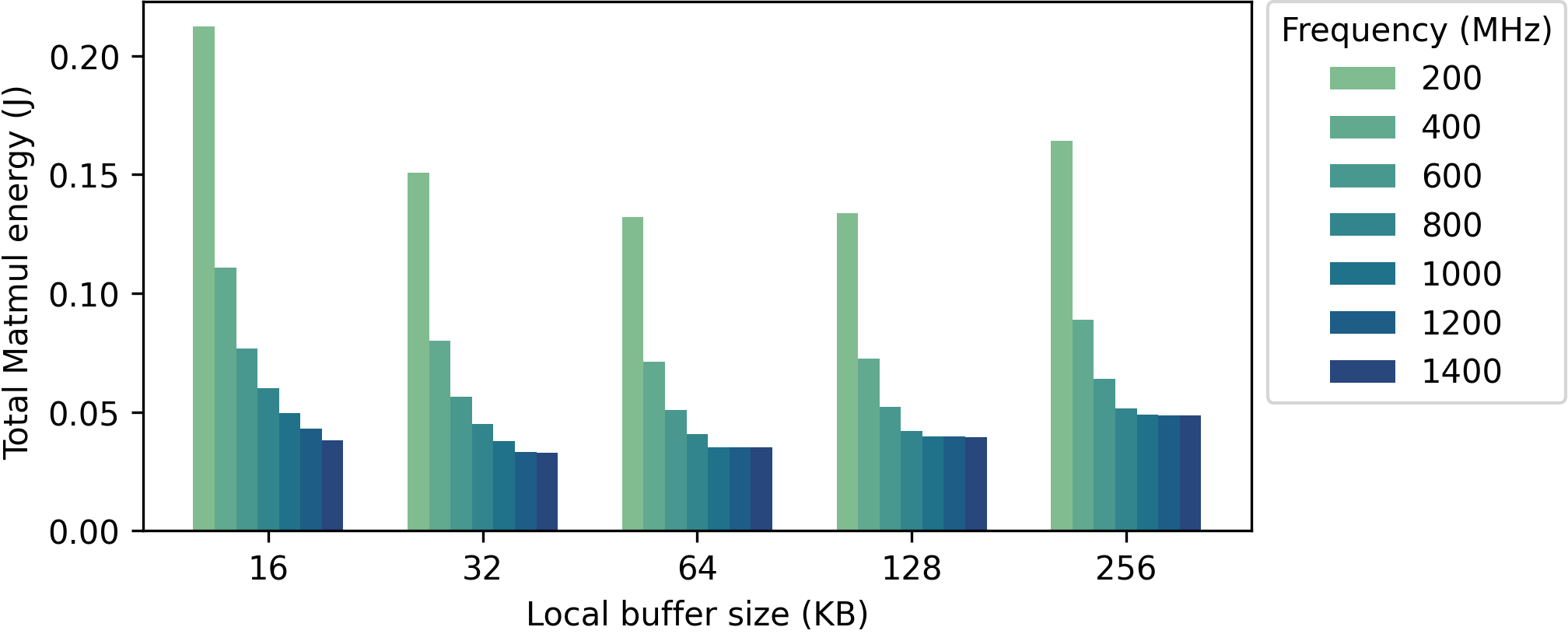}
    \caption{Decode Total Energy Quadruple Memory Bandwidth}
    \label{decode_total_energy_quadruple_memory_bandwidth}
  \end{subfigure}

  \begin{subfigure}{0.32\textwidth}
    \centering
    \includegraphics[width=\linewidth]{images/energy_delay_product_decode_edp_baseline_bandwidth.png}
    \caption{Energy Delay Product Baseline Memory Bandwidth}
    \label{energy_delay_product_baseline_memory_bandwidth}
  \end{subfigure}
      \hfill
  \begin{subfigure}{0.32\textwidth}
    \centering
    \includegraphics[width=\linewidth]{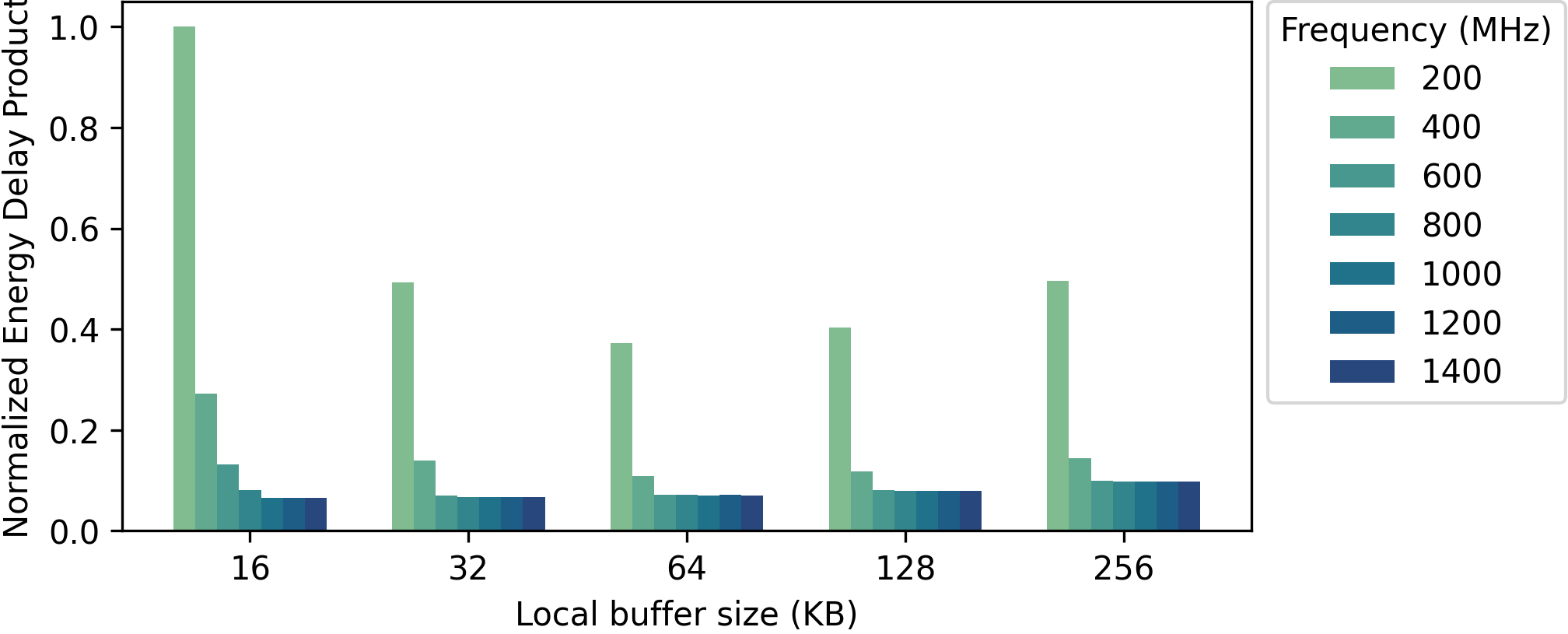}
    \caption{Energy Delay Product Double Memory Bandwidth}
    \label{energy_delay_product_double_memory_bandwidth}
  \end{subfigure}
      \hfill
  \begin{subfigure}{0.32\textwidth}
    \centering
    \includegraphics[width=\linewidth]{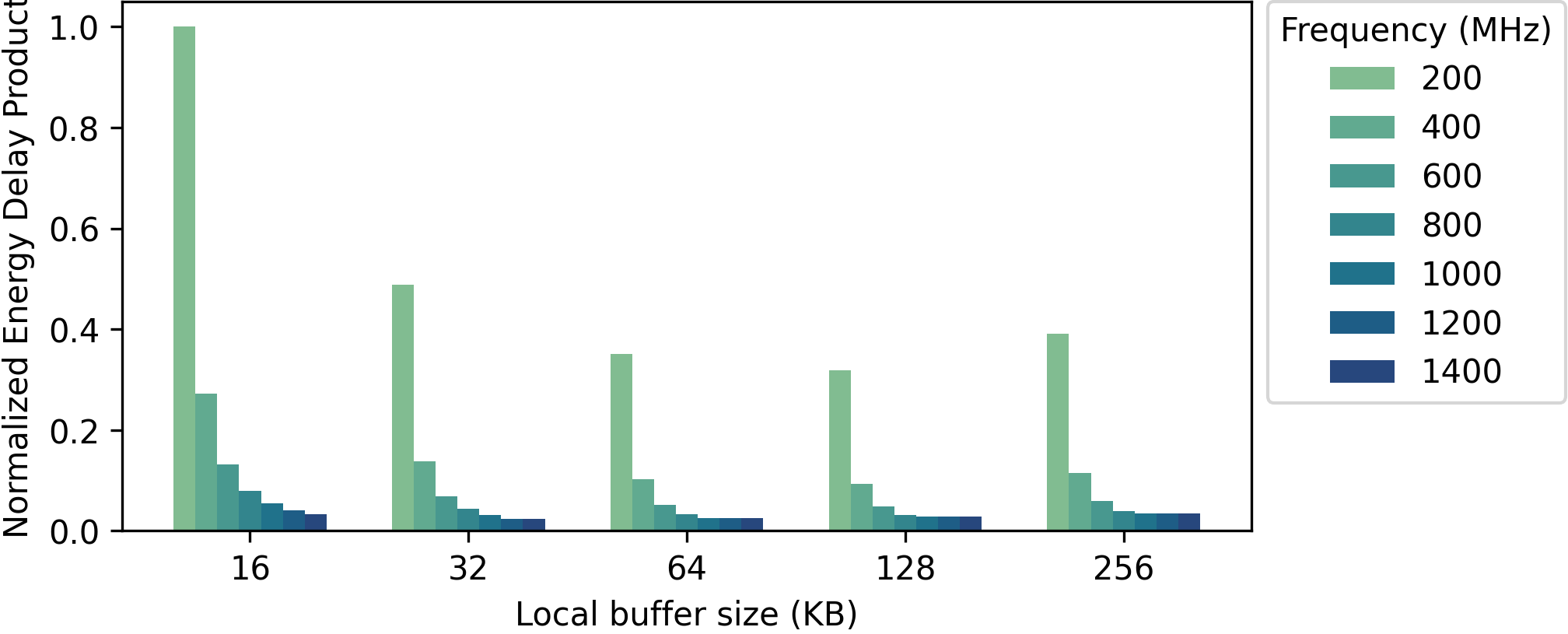}
    \caption{Energy Delay Product Quadruple Memory Bandwidth}
    \label{energy_delay_product_quadruple_memory_bandwidth}
  \end{subfigure}

  \caption{Comparison of Performance roofline, Latency, Energy, and EDP for Decode as a function of Memory Bandwidth. First (left) graph column : 2048 GB/s, second (middle) graph column: 4096 GB/s, third (right) graph column: 8192 GB/s.}
  \label{fig:grid6_mem_bandwidth}
\end{figure*}

%%%%%%% Effects of increasing memory bandwidth %%%%%%%%%
\subsection{Effect of Increasing Memory Bandwidth}
Decode is heavily affected by memory bandwidth, because decode latency is primarily a result of waiting on memory. Figures \ref{roofline_baseline_memory_bandwidth} and \ref{roofline_double_memory_bandwidth} show the effect of doubling the baseline memory of 2048 GB/s to 4096 GB/s, and Figure~\ref{roofline_quadruple_memory_bandwidth} shows the effect of quadrupling the memory bandwidth to 8192 GB/s. Increasing memory bandwidth raises the roofline of the performance curves, allowing higher frequencies to improve performance before becoming memory-bound. However, performance gains from increased bandwidth show diminishing returns; for example, quadrupling bandwidth only resulted in approximately a threefold increase in the roofline level. Figures \ref{decode_latency_baseline_memory_bandwidth}, \ref{latency_decode_double_memory_bandwidth}, and \ref{latency_decode_quadruple_memory_bandwidth}, show the effect on decode latency from doubling and quadrupling memory bandwidth. Figures \ref{decode_total_energy_baseline_memory_bandwidth}, \ref{decode_total_energy_double_memory_bandwidth}, and \ref{decode_total_energy_quadruple_memory_bandwidth}, show the effect on decode total energy from doubling and quadrupling memory bandwidth. At the baseline memory bandwidth, a 32KB local buffer is ideal for the lowest frequencies, while doubling the memory bandwidth makes a 64KB local buffer preferred. Figures \ref{energy_delay_product_baseline_memory_bandwidth}, \ref{energy_delay_product_double_memory_bandwidth},  and \ref{energy_delay_product_quadruple_memory_bandwidth}, show the effect on decode energy delay product from doubling and quadrupling memory bandwidth. At the baseline memory bandwidth, the ideal $S =$ 32KB and the ideal $f =$ 600MHz. Quadrupling memory bandwidth results in the lowest EDP at $S =$ 128KB and an ideal $f =$ 1000MHz. The results show that quadrupling memory bandwidth decreases both latency and total energy consumption during decode, and that increasing memory bandwidth affects the ideal $S$ and $f$ to minimize EDP. 

%\subsection{Overview }
\subsection{Main Findings Summary}
\label{sec:main_findings}

%\begin{itemize}
  %\item 
  %\textbf{Prefill is compute-bound; decode is bandwidth-bound.}
  It is well known that Prefill is compute-bound and  decode is bandwidth-bound. Prefill latency is dominated by compute and improves with higher operating frequency, while decode spends the majority of time stalled on memory (compute is a small fraction of decode latency) and becomes largely insensitive to further frequency scaling once memory-bound.

  Our work refines this collective knowledge with the following conclusions:

  %\item 
  \textbf{Local SRAM size exhibits strong diminishing returns beyond 32\,KB.}
  Increasing the local buffer from 16\,KB to 32\,KB yields a clear latency benefit, but larger buffers (64\,KB and above) provide progressively smaller additional reductions in latency for both phases.

  %\item 
  \textbf{Total energy is predominantly driven by SRAM size via leakage.}
  Larger SRAMs significantly increase static energy (leakage) and increase dynamic energy due to higher capacitive switching in larger arrays; these costs are not generally offset by proportional latency improvements, especially beyond 32--64\,KB.

  %\item 
  \textbf{Counter-intuitively, higher frequency can reduce total energy.}
  Although dynamic power increases with frequency, the reduced execution time can lower static energy enough that total energy decreases within the explored frequency range, particularly for the compute-bound prefill phase.

  %\item 
  \textbf{Best energy/EDP occurs at small SRAM and high frequency for the baseline system.}
  Across the evaluated design points, configurations with \mbox{32--64\,KB} local SRAM and high frequency (approximately \mbox{1200--1400\,MHz}) provide the best overall balance of latency and energy efficiency (lowest EDP); decode energy is most favorable at \mbox{32\,KB}.

  %\item 
  \textbf{Decode scalability is capped by memory bandwidth; bandwidth shifts the optimum.}
  Roofline analysis explains decode behavior: performance improves with frequency only until the memory-bandwidth ceiling is reached. Increasing memory bandwidth raises this ceiling (with diminishing returns), and it can shift the EDP-optimal point toward larger SRAM sizes and different frequencies (e.g., higher bandwidth making 64\,KB or 128\,KB preferable in decode).

%%%%%%%%%%%%%%%%%%%% OLD CONCLUSION

\section{Conclusion}

This paper quantifies how on-chip SRAM capacity and  frequency jointly shape the performance and energy efficiency of LLM inference, explicitly separating the compute-bound prefill phase from the memory-bound decode phase. Using a combined simulation workflow, OpenRAM for SRAM power/energy modeling, LLMCompass for matmul-layer latency, and ScaleSIM for systolic array utilization and data transfer activity, plus post-layout systolic-array estimates for utilization and activity—we evaluated an accelerator-style design across local-buffer sizes (16KB–256KB), frequencies (200–1400MHz), and multiple external-memory bandwidth points. 

Across both phases, we find that total energy is dominated by SRAM size, primarily because larger buffers impose a leakage “tax” (and higher per-access switching energy) that is rarely repaid by proportional latency reductions. Latency benefits from increasing the local buffer show strong diminishing returns beyond $\approx$32KB (and modestly to 64KB), while energy continues to rise with larger SRAMs. 

Frequency scaling has an asymmetric impact. Prefill latency improves meaningfully with higher operating frequency because it is compute-limited, and the reduced runtime can lower total energy by cutting static energy more than the added dynamic power---making high frequency attractive within the explored range. In decode, performance quickly encounters a memory-bandwidth ceiling: once memory-bound, increasing frequency yields little to no latency reduction, so energy trends are largely set by SRAM size rather than compute speed. Roofline-style analysis and bandwidth sweeps confirm that raising memory bandwidth increases decode’s performance ceiling (with diminishing returns) and can shift the EDP-optimal operating point toward larger buffers and different frequencies. 

For the baseline system studied, the best overall energy–performance trade-off occurs with small local buffers (32KB--64KB) and high frequency ($\approx$ 1200 to 1400MHz), with decode energy typically favoring 32KB. More broadly, the results imply that designing energy-efficient LLM accelerators should prioritize (1) avoiding oversized on-chip SRAMs that inflate leakage, and (2) treating decode optimization as a bandwidth problem rather than a pure compute-frequency problem—since frequency gains saturate once the workload becomes memory-bound.


\begin{thebibliography}{00}

\bibitem{iea}
IEA, ``Energy and AI,'' \textit{International Energy Agency Report}, 2025. [Online]. Available: \url{https://www.iea.org/reports/energy-and-ai}

\bibitem{rado}
R. Desislavov, F. Martínez-Plumed, and J. Hernández-Orallo, ``Trends in AI Inference Energy Consumption: Beyond the Performance-\textit{vs}-Parameter Laws of Deep Learning,'' \textit{Sustainable Computing: Informatics and Systems}, 2023.

\bibitem{efficient}
C. Niu, W. Zhang, Y. Zhao, and Y. Chen, ``Energy Efficient or Exhaustive? Benchmarking Power Consumption of LLM Inference Engines,'' \textit{ACM SIGEnergy Energy Informatics Review}, vol.~5, no.~2, pp.~56--62, Jul.~2025.

\bibitem{npu}
X. Yuqi, J. Huang. ``ReGate: Enabling Power Gating in Neural Processing Units,'' in \textit{International Symposium on Microarchitecture (MICRO)}, 2025.

\bibitem{quantifying}
M. Özcan, P. Wiesner, P. Weiß, and O. Kao, ``Quantifying the Energy Consumption and Carbon Emissions of LLM Inference via Simulations,'' \textit{arXiv:2507.11417}, Jul. 2025. [Online]. Available: http://arxiv.org/abs/2507.11417

\bibitem{ellie}
H. Fan, Y.-C. Lin, and V. Prasanna, ``ELLIE: Energy-Efficient LLM Inference at the Edge Via Prefill-Decode Splitting,'' in \textit{International Conference on Application-specific Systems, Architectures and Processors (ASAP)}, 2025.

\bibitem{MNN}
Z. Huang \textit{et al.}, ``MNN-AECS: Energy Optimization for LLM Decoding on Mobile Devices via Adaptive Core Selection,'' \textit{arXiv:2506.19884}, Jun. 2025. [Online]. Available: http://arxiv.org/abs/2506.19884

\bibitem{hungry}
N. Jegham, M. Abdelatti, L. Elmoubarki, and A. Hendawi, ``How Hungry is AI? Benchmarking Energy, Water, and Carbon Footprint of LLM Inference,'' \textit{arXiv:2505.09598}, Jul. 2025. [Online]. Available: http://arxiv.org/abs/2505.09598

\bibitem{guthaus2016openram}
M. R. Guthaus, J. E. Stine, S. Ataei, B. Chen, B. Wu, M. Sarwar, ``OpenRAM: An Open-Source Memory Compiler,'' in \textit{International Conference on Computer-Aided Design (ICCAD)}, 2016.

\bibitem{zhang2024llmcompass}
H. Zhang, August Ning. R. B. Prabhakar, D. Wentzlaff, ``LLMCompass: Enabling Efficient Hardware Design for Large Language Model Inference,'' in \textit{International Symposium on Computer Architecture (ISCA)}, 2024.

\bibitem{raj2025scalesim}
R. Raj, S. Banerjee, N. Chandra,  Z. Wan, J. Tong, A. Samajdar, T. Krishna, ``SCALE-Sim v3: A Modular Cycle-Accurate Systolic Accelerator Simulator for End-to-End System Analysis,'' in \textit{International Symposium on Performance Analysis of Systems and Software (ISPASS)}, 2025.

\bibitem{tpu}
N. P. Jouppi, C. Young, N. Patil \textit{et al.}, ``In-Datacenter Performance Analysis of a Tensor Processing Unit,'' in \textit{International Symposium on Computer Architecture (ISCA)}, 2017.

\bibitem{purdue}
H. Li, M. Bhargav, P. N. Whatmough and H.-S. P. Wong, ``On-Chip Memory Technology Design Space Explorations for Mobile Deep Neural Network Accelerators,'' in \textit{Design Automation Conference (DAC)}, 2019.

\bibitem{yosys}
C.~Wolf, J.~Glaser, and J.~Kepler,
``Yosys---A Free Verilog Synthesis Suite,''
in \textit{Austrochip Workshop on Microelectronics}, 2013.

\bibitem{openroad_1}
A. B.~Kahng and T.~Spyrou, ``The OpenROAD Project: Unleashing Hardware Innovation'', in \textit{Government Microcircuit Applications and Critical Technology Conference}, 2021. 

\bibitem{freepdk}
NCSU Electronic Design Automation Group,
\emph{FreePDK45: An Open-Source Predictive Process Design Kit}, North Carolina State University, 2013. [Online]. Available: http://www.eda.ncsu.edu/wiki/FreePDK45

\bibitem{dao2023flashattention2fasterattentionbetter}
T. Dao, ``FlashAttention-2: Faster Attention with Better Parallelism and Work Partitioning,'' \textit{arXiv:2307.08691}, 2023. [Online]. Available: https://arxiv.org/abs/2307.08691

\bibitem{Keramidas}
G. Keramidas, V. Spiliopoulos, and S. Kaxiras, ``Interval-Based Models for Run-Time DVFS Orchestration in SuperScalar Processors,'' in \textit{International Conference on Computing Frontiers (CF)}, 2010.

\bibitem{Lieven}
S. Eyerman and L. Eeckhout, ``Fine-Grained DVFS Using On-Chip Regulators,'' \textit{ACM Transactions on Architecture and Code Optimization (TACO)}, vol.~8, no.~1, pp.~1--24, Feb. 2011.

\bibitem{Kaxiras_Martonosi_2008}
S. Kaxiras and M. Martonosi, \textit{Computer Architecture Techniques for Power-Efficiency}. Morgan \& Claypool, 2008.

\bibitem{Spiliopoulos}
V.~Spiliopoulos, A.~Sembrant, G.~Keramidas, and E.~Hagersten, ``A Unified DVFS-Cache Resizing Framework,'' Technical Report, Department of Information Technology, Uppsala University, 2016.
[Online]. Available: \url{urn:nbn:se:uu:diva-300840}

\bibitem{IJECCE297}
S. A. Dwivedi, ``Study of Minimization of Power Dissipation Techniques Used in SRAM Cell,'' \textit{International Journal of Electronics Communication and Computer Engineering (IJECCE)}, vol.~3, no.~2, pp.~368--371, 2012.

\bibitem{openroad_2}
T. Ajayi, V. A. Chhabria, M. Fogaça, \textit{et al.}, ``Toward an Open-Source Digital Flow: First Learnings from the OpenROAD Project,'' in \textit{Design Automation Conference (DAC)}, 2019.

\bibitem{kaxiras2001cache}
S.~Kaxiras, Z.~Hu, and M.~Martonosi,
``Cache Decay: Exploiting Generational Behavior to Reduce Cache Leakage Power,'' in \textit{International Symposium on Computer Architecture (ISCA)}, 2001.

\bibitem{flautner2002drowsy}
K.~Flautner, N.~S.~Kim, S.~Martin, D.~Blaauw, and T.~Mudge, ``Drowsy Caches: Simple Techniques for Reducing Leakage Power,'' \textit{ACM SIGARCH Computer Architecture News}, vol.~30, no.~2, pp.~148--157,~2002.

% @inproceedings{kaxiras2001cache,
%   title={Cache decay: Exploiting generational behavior to reduce cache leakage power},
%   author={Kaxiras, Stefanos and Hu, Zhigang and Martonosi, Margaret},
%   booktitle={Proceedings of the 28th annual international symposium on Computer architecture},
%   pages={240--251},
%   year={2001}
% }

% @article{flautner2002drowsy,
%   title={Drowsy caches: simple techniques for reducing leakage power},
%   author={Flautner, Kriszti{\'a}n and Kim, Nam Sung and Martin, Steve and Blaauw, David and Mudge, Trevor},
%   journal={ACM SIGARCH Computer architecture news},
%   volume={30},
%   number={2},
%   pages={148--157},
%   year={2002},
%   publisher={ACM New York, NY, USA}
% }

\end{thebibliography}
\end{document}